\documentclass[useAMS,usenatbib]{mn2e}
\usepackage{astrojournals,amsmath,graphicx,txfonts,fleqn}
\usepackage[normalem]{ulem}

\newcommand*{\diff}       {\mathop{}\!\mathrm{d}}
 
\newcommand*{\Msun}       {\mbox{M$_\odot$}}
\newcommand*{\Robs}       {R_{\mathrm{p}}}
\newcommand*{\rperi}      {r_{\mathrm{peri}}}
\newcommand*{\rapo}       {r_{\mathrm{apo}}}
\newcommand {\corr}[1]    {#1}

\title[The mass distribution of the Fornax dSph]
      {The mass distribution of the Fornax dSph:\\
        constraints from its globular cluster distribution}

\author[Cole, Dehnen, Read \& Wilkinson]
       {David R.~Cole$^1$, Walter Dehnen$^1$,
         Justin I.~Read$^{1,2}$, Mark I.~Wilkinson$^1$\thanks{
           Email: david.cole@le.ac.uk, walter.dehnen@le.ac.uk,
           jread@phys.ethz.ch, mark.wilkinson@le.ac.uk}\\
  $^1$Department of Physics \& Astronomy,
  University of Leicester,
  Leicester, LE1~7RH\\
  $^2$Institute for Astronomy, Department of Physics,
  ETH Z\"urich, Wolfgang-Pauli-Strasse 16, CH-8093 Z\"urich, Switzerland
}
\date{Accepted .
      Received ;
      }
\pagerange{\pageref{firstpage}--\pageref{lastpage}}
\pubyear{2012}
\begin{document}
\maketitle
\label{firstpage}
\begin{abstract}
  Uniquely among \corr{the dwarf spheroidal (dSph)} satellite \corr{galaxies}
  of the Milky Way, Fornax \corr{hosts} globular clusters. It remains a puzzle
  as to why dynamical friction has not yet dragged any of \corr{Fornax's five
    globular clusters} to the centre, and also why there is no evidence that
  any similar star cluster has been \corr{in the past} (\corr{for Fornax or}
  any other dSph). We set up a suite of \corr{2800} $N$-body simulations that
  sample the \corr{full} range of globular-cluster \corr{orbits} and mass
  models consistent with all existing observational constraints \corr{for
    Fornax}. In agreement with previous work, we find that if Fornax has a
  large dark-matter core then its globular clusters remain close to their
  currently observed locations for long times. Furthermore, we find previously
  unreported behaviour for clusters that start inside the core region. These
  are pushed out of the core and gain orbital energy, a process we call
  `dynamical buoyancy'.  Thus a cored mass distribution in Fornax will
  naturally lead to a shell-like globular cluster distribution near the core
  radius, independent of the initial conditions. By contrast, CDM-type cusped
  mass distributions lead to the rapid infall of at least one cluster within
  \corr{$\Delta t=1$}-2\,Gyr\corr{, except when picking unlikely initial
    conditions for the cluster orbits ($\sim2\%$ probability), and almost all
    clusters within \corr{$\Delta t=10$}\,Gyr}. Alternatively, if Fornax has
  only a weakly cusped mass distribution, dynamical friction is much
  reduced. While over $\corr{\Delta} t=10$\,Gyr this still leads to the infall
  of 1-\corr{4} clusters from their present orbits, the infall of any cluster
  within $\corr{\Delta} t=1$-2\,Gyr is much less likely (with probability
  0-70\%, depending on $\corr{\Delta} t$ and the strength of the cusp). Such a
  solution to the timing problem requires (in addition to a shallow
  dark-matter cusp) that in the past the globular clusters were somewhat
  further from Fornax than today; they most likely did not form within Fornax,
  but were accreted.
\end{abstract}
\begin{keywords}
  stellar dynamics -- 
  methods: $N$-body simulations -- 
  galaxies: kinematics and dynamics --
  galaxies: structure --
  galaxies: haloes
\end{keywords}

%%%%%%%%%%%%%%%%%%%%%%%%%%%%%%%%%%%%%%%%%%%%%%%%%%%%%%%%%%%%%%%%%%%%%%%%%%%%%%%%
\section{Introduction}
\label{sec:intro}
The Fornax dwarf spheroidal (dSph) galaxy is the most massive undisrupted dSph
satellite of the Milky Way \citep{WalkerEtAl2009}. Like all dSphs it is dark
matter dominated even in its central regions. It is unique among the
undisrupted dSphs in having globular clusters; it has five, with three of them
at a projected distance outside of the half light radius (see table
\ref{tab:GCdata}). There is also evidence of two shell-like structures, which
may be the remnants of a merger occurring more than 2 Gyr ago
\citep{ColemanEtAl2004,ColemanEtAl2005}.

One apparent paradox about these clusters is that, because they \corr{appear
  to orbit in a} massive background of dark matter, they should be affected by
dynamical friction which will cause their orbits to decay. Fornax's globular
clusters are metal poor and very old, comparable to the oldest globular
clusters in the Milky Way with ages of the order of a Hubble time
\citep{BuonannoEtAl1998, BuonannoEtAl1999, MackeyGilmore2003b,
  GrecoEtAl2007}. During their lifetime it would be expected that they fall to
the centre of Fornax and form a nuclear star cluster
\citep*{TremaineOstrikerSpitzer1975, Tremaine1976}. However, no bright stellar
nucleus is observed in Fornax, or in fact any other dSph. This is known as the
timing problem for Fornax's clusters because it seems highly improbable that
Fornax's globular clusters would be observed just briefly before they fall
into the core.

Several solutions to the timing problem have been proposed.
\cite*{OhLinRicher2000} suggested two ideas. First, that a population of black
holes transferred energy to the clusters through close encounters; and second,
that a strong tidal interaction between the Milky Way and Fornax could inject
energy into their orbits. There is no observational evidence for a population
of black holes in the centre of Fornax, while the currently observed proper
motion indicates that the orbit of Fornax around the Milky Way never takes it
closer than at present (\citealt{DinescuEtAl2004};
\citealt*{LuxReadLake2010}). Thus, both ideas appear to be
disfavoured. \cite{AngusDiaferio2009} proposed that all but the most massive
cluster could avoid sinking into the centre of Fornax if their current
distance is much larger than projected but still within the tidal radius of
Fornax of $\sim1.9\,$kpc. However, this (i) requires special arrangement of
the current projected positions and (ii) stills leaves a timing problem for
the most massive cluster and is therefore not a complete solution (moreover,
their analysis was based on Chandrasekhar's simple dynamical-friction formula,
which is not suitable for accurate estimates).

Using numerical simulations and analytic arguments \cite{GoerdtEtAl2006}
proposed that the current distribution of the Fornax clusters can be explained
by the diminution of dynamical friction on the edge of a cored matter
distribution which causes the clusters to remain outside the dark matter core
radius. Dynamical reasons for this `core-stalling' effect have been explored
in \cite{ReadEtAl2006a}, \cite{Inoue2009}, and
\cite*{ColeDehnenWilkinson2011}. Support for this result was provided by
\cite*{Sanchez-SalcedoReyes-IturbideHernandez2006} who showed that a cored
matter distribution in dwarf galaxies can significantly delay the infall times
of the globular clusters (even if Chandrasekhar's simple dynamical-friction
formula is used).

Measuring and/or constraining the dark-matter distribution in Fornax is
interesting as a test of our current cosmological model. Collisionless
cosmological simulations (which ignore the effects of baryons) predict a
universal density distribution for dark matter halos, with a central density
cusp $\rho\propto r^{-\gamma}$ where $\gamma\sim1$
\citep*{DubinskiCarlberg1991, NavarroFrenkWhite1996}. If the dark matter
distribution in Fornax is found to deviate strongly from this prediction, this
could imply that baryons have an important dynamical role in shaping the
central dark matter distribution in dwarf galaxies
(e.g.\ \citealt*{NavarroEkeFrenk1996}, \citealt*{ElZantShlosmanHoffman2001},
\citealt{ReadGilmore2005}, \citealt{GoerdtEtAl2010},
\citealt{ColeDehnenWilkinson2011}, \citealt{PontzenGovernato2012}), or that we
must turn to more exotic cosmological models
(e.g.\ \citealt{TremaineGunn1979}, \citealt{KochanekWhite2000},
\citealt{HoganDalcanton2000}, \citealt{StrigariEtAl2006},
\citealt{Villaescusa-NavarroDalal2011}, \citealt{MaccioEtAl2012}).

The first evidence that dSphs may have a constant density core came from
\cite*{KleynaWilkinsonGilmore2003} who found indirect evidence for a core in
the Ursa Minor dSph. The Milky Way dSphs have been observed intensively in
recent years, primarily because these systems are the most dark-matter
dominated known. They contain mostly intermediate or old stellar populations
which are likely to be well mixed in the dark matter potential because star
formation ceased many dynamical times ago. This in turn implies that they are
ideal laboratories for studying the mass structure of their dark matter
halos. The intense observational effort means that there is a wealth of
kinematical data available to form the basis for theoretical models of these
systems. One line of approach has been based on the Jeans equations where a
parametric light profile for the stars is assumed and a velocity dispersion
profile is derived based on a underlying parameterised dark matter profile
(\citealt*{PenarrubiaMcConnachieNavarro2008}, \citealt{StrigariEtAl2008},
\citealt{WalkerEtAl2009}). This approach has its drawbacks (see for example
\citealt{AmoriscoEvans2011}). Most importantly, the degeneracy between the
mass profile and the velocity anisotropy (which is poorly constrained), means
that both cusped and cored density distributions are consistent with even the
latest data. However, modelling the dSphs as two chemically distinct
populations with different scale lengths, it appears that this degeneracy can
be broken \citep{BattagliaEtAl2008, AmoriscoEvans2011b, AmoriscoEvans2011,
  WalkerPenarrubia2011}; the results favour a cored density distribution in
the two dSph galaxies best analysed to date: Fornax and Sculptor \citep[but
  see][]{BreddelsEtAl2012}.

In this paper, we we follow the work of \cite{GoerdtEtAl2006} by examining
what the current location of the globular clusters can tell us about Fornax's
mass distribution. Our work improves on this previous analysis in several key
respects: (i) we use several mass models for the underlying potential in
Fornax that sample the full range consistent with the latest data; (ii) we use
the latest data for Fornax's globular clusters as constraints on their phase
space distribution; and (iii) we run thousands of $N$-body models to sample
the uncertainties in the cluster distribution and Fornax mass model.

This large search of the available parameter space allows us to address
whether or not there are multiple solutions to Fornax's timing problem.  To
focus the discussion, we phrase the timing problem as a contradiction with
either of the following two hypotheses
\begin{enumerate}
\item \label{hyp:long} The Fornax globular cluster system is in a near-steady
  state, consequently none of the globular clusters should fall into the core
  of Fornax within a Hubble time.
\item \label{hyp:short} Our present cosmic epoch of observing Fornax is not
  special, consequently the system does not evolve significantly on a time
  scale short compared to a Hubble time. In particular, within 1-2\,Gyr none
  of the clusters should fall into the core of Fornax with high probability.
\end{enumerate}
The first hypothesis can be justified by the assumption that the present state
of Fornax's globular cluster system is representative also for its past. This
assumption, however, is not necessary, as one can easily think of alternative
scenarios. The second hypothesis, on the other hand, is harder to avoid and
similar to the cosmological principle, though here expressed in terms of the
epoch of observation rather than its vantage point and orientation.  We will
refer to contradictions with these hypotheses as the \emph{long-term} and
\emph{immediate} timing problem, respectively.

This paper is organised as follows.
Section~\ref{sec:fornax} reviews the properties of the Fornax system relevant
for our study,
Section~\ref{sec:model} details our modelling approach,
Sections~\ref{sec:results}\&\ref{sec:sinking} present the simulations results
and assess the probability that none of the five clusters will sink into the
core of Fornax within either 1--2\,Gyr or a Hubble time.
Finally, in Section~\ref{sec:conclusions} we discuss the implications of our
results and draw our conclusions.

%%%%%%%%%%%%%%%%%%%%%%%%%%%%%%%%%%%%%%%%%%%%%%%%%%%%%%%%%%%%%%%%%%%%%%%%%%%%%%%%
%
\begin{table}
  \setlength{\tabcolsep}{0.48em}
  %\begin{minipage}{84mm}
    %\begin{center}
      \begin{tabular}{@{}l r@{$\;\!\pm\;\!$}l c c c r@{$\;\!\pm\;\!$}l r@{$\;\!\pm\;\!$}l@{}} 
      Object &
      \multicolumn{2}{c}{log$M$ $^a$} &
      \multicolumn{1}{c}{$M$ $^a$} &
      \multicolumn{1}{c}{$r_{\mathrm{c}}$ $^a$} &
      \multicolumn{1}{c}{$\Robs$ $^a$} &
      \multicolumn{2}{c}{$d_{\mathrm{los}}$} &
      \multicolumn{2}{c}{$\Delta\upsilon_{\mathrm{los}}$$^d$} \\
      & 
      \multicolumn{2}{c}{$[\Msun]$} & 
      [$10^5\Msun$] 
      & [pc] & [kpc] & 
      \multicolumn{2}{c}{[kpc]} & 
      \multicolumn{2}{c}{[km\,s$^{-1}$]} \\
      \hline
      % GC log Mgc Mgc  rc  rp  rlos vlos  
      dSph &
      \multicolumn{2}{c}{$8.15^{\,+\,0.19}_{\,-\,0.37}{}^e$} &
      1420\,$^e$ & 668$^e$ & - & 
      %\multicolumn{2}{c}{$>10$ ${}^i$} &
      137\phantom{.0} & 13\phantom{.}$^{b,f}$ & \multicolumn{2}{c}{-}\\
      & \multicolumn{2}{c}{} & & & &
      138\phantom{.0} & 8\phantom{.0}$^{g}$ & \multicolumn{2}{c}{} \\
      %\hline
      % GC log Mgc Mgc  rc  rp  rlos vlos  
      GC1 & 4.57&0.13 & 0.37 & 10.03 & 1.6\phantom{1} & 130.6&3.0$^{b}$ & \multicolumn{2}{c}{-} \\
      GC2 & 5.26&0.12 & 1.82 & \phantom{1}5.81 & 1.05 & 136.1&3.1$^{b}$ & --1.2&4.6 \\
      GC3 & 5.56&0.12 & 3.63 & \phantom{1}1.60 & 0.43 & 135.5&3.1$^{b}$ & 7.1&3.9 \\
      GC4 & 5.12&0.24 & 1.32 & \phantom{1}1.75 & 0.24 & 134.0&6\phantom{.0}$^{c}$ & 5.9&3.4 \\
      GC5 & 5.25&0.20 & 1.78 & \phantom{1}1.38 & 1.43 & 140.6&3.2$^{b}$ & 8.7&3.6 \\
    \end{tabular}
    \caption{Data for the Fornax system. Column 3 shows the most likely mass
      from column 2. $r_{\mathrm{c}}$ is the 
      King (1962)%\cite{King1962} 
      model core radius
      for the globular clusters, and the half-light radius for the
      dSph. $R_{\mathrm{p}}$ is the projected distance of the cluster from the
      centre of Fornax. $d_{\mathrm{los}}$ is the distance to each cluster and
      $\Delta\upsilon_{\mathrm{los}}$ the line-of-sight velocity relative to
      Fornax itself.  References: 
      $^a$Mackey \& Gilmore (2003b),%\cite{MackeyGilmore2003a},
      $^b$Mackey \& Gilmore (2003a),%\cite{MackeyGilmore2003b}, 
      $^c$Greco et al. (2007),%\cite{GrecoEtAl2007},
      $^d$Mateo et al. (1991),%\cite{MateoOlszewskiWelchFischerKunkel1991},
      $^e$Walker et al. (2009),%\cite{WalkerEtAl2009}, 
      $^f$Buonanno et al. (1999),%\cite{BuonannoEtAl1999},
      $^g$Mateo (1998),%\cite{Mateo1998}, 
      $^h$Buonanno et al. (1998).%\cite{BuonannoEtAl1998}.
      \label{tab:GCdata}
    }
    %\end{center}
  %\end{minipage}
\end{table}
\section{The Fornax system}
\label{sec:fornax}
We summarise in Table~\ref{tab:GCdata} the most relevant data for our study
and their origin.
%%%%%%%%%%%%%%%%%%%%%%%%%%%%%%%%%%%%%%%%%%%%%%%%%%%%%%%%%%%%%%%%%%%%%%%%%%%%%%%%
\subsection{The dSph}
\label{sec:fornax:dSph}
As already discussed above, Fornax is the most massive of the Milky Way's dSph
(except possibly for disrupted objects such as Sgr dwarf) and unique amongst
(undisrupted) dSph to host a globular cluster system. The very fact that
Fornax can hold onto a globular-cluster system implies that Galactic tides
cannot be strong enough to pull these clusters off. This in turn requires that
its Galactic orbit never carries Fornax too close to the Milky Way, where the
tidal forces become exceedingly strong.

\cite*{LuxReadLake2010} estimate the peri-galactic radius of Fornax, based on
its observed position, distance, radial velocity and proper motion, to be
100--130\,kpc, which indeed is only slightly smaller than its current distance
of about 140\,kpc (see Table~\ref{tab:GCdata}).

Based on this result, we estimate the tidal radius for Fornax using the method
of \cite{ReadEtAl2006b}, where the dSph is modelled as a spherical satellite
orbiting the Milky Way represented by a \cite{Hernquist1990} model. We then
solve equation~(7) of \cite{ReadEtAl2006b}, which accounts for the orbit of
the satellite about the host, and the orbits of the stars within the
satellite. We find a tidal radius of 1.8--2.8\,kpc, based on a the range of
masses for Fornax given in Table~\ref{tab:mass:models} and using the extremal
values for the the orbital data taken from \cite{LuxReadLake2010} and a total
(extended) mass for the Milky Way of $1{-}2\times10^{12}\,\Msun$.
%%%%%%%%%%%%%%%%%%%%%%%%%%%%%%%%%%%%%%%%%%%%%%%%%%%%%%%%%%%%%%%%%%%%%%%%%%%%%%%%
\subsection{The globular clusters}
\label{sec:fornax:GCs}
Our principle sources for globular-cluster data are those published by
\cite{MackeyGilmore2003b,MackeyGilmore2003a} and \cite{GrecoEtAl2007} who have
carried out thorough surveys of the Fornax globular clusters. For our purposes
the main data required are the clusters' masses, sizes, three dimensional
positions and velocities.

The best estimates for these quantities are given in
Table~\ref{tab:GCdata}. The values for the core radius $r_{\mathrm{c}}$ of
each cluster are based on the surface brightness profiles calculated in
\cite{MackeyGilmore2003a}. These are \citeauthor*{ElsonFallFreeman1987} (1987,
EEF) models and the \cite{King1962} model core radius $r_{\mathrm{c}}$ is
related to the EFF scale parameter $a$ by $r_{\mathrm{c}}=a(2^{2/\gamma} - 1
)^{1/2}$, where $\gamma$ is the power law slope of the surface brightness at
large radii.

\begin{figure}
  \centerline{
    \resizebox{70mm}{!}{\includegraphics{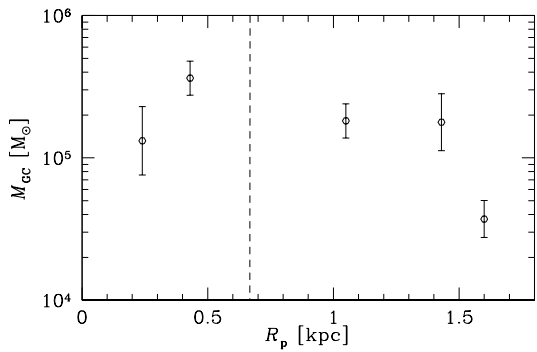}}
  }
  \vspace*{-0.3ex}
  \caption{
    \label{fig:GC:R:M}
    Distribution of the Fornax globular clusters in mass and projected
    distance from the centre of the Fornax dSph. The dashed vertical line
    indicates the stellar half-light radius of the dSph.  }
\end{figure}
Fig.~\ref{fig:GC:R:M} plots the distribution of the five Fornax globular
clusters in mass $M$ and projected radius $\Robs$ from the dSph. There are two
interesting observations to be made from this figure. First, the radial
distribution of clusters is consistent with that of the stars within Fornax:
there is about half of the total cluster light within the stellar half-light
radius of 668\,pc. There are two possible interpretations of this. Either it
is a coincidence, or the formation histories of the clusters and the Fornax
galaxy are closely related, in particular, the clusters formed within the same
entities as the stars.

Second, there is a weak correlation between $M$ and $\Robs$. In particular,
the lightest cluster is furthest away from Fornax and the heaviest is the
second closest. The remaining three are about equally massive and cover a
spread of $\Robs$. This correlation is in the sense expected from mass
segregation such as driven by dynamical friction.

%%%%%%%%%%%%%%%%%%%%%%%%%%%%%%%%%%%%%%%%%%%%%%%%%%%%%%%%%%%%%%%%%%%%%%%%%%%%%%%%
\section{Modelling Approach}
\label{sec:model}
The basis for our approach is to take the most up-to-date observations of
Fornax's globular clusters and combine these with plausible mass models
consistent with the latest kinematic data for Fornax's stars. We then create,
for each Fornax mass model, initial conditions for the Fornax globular-cluster
system, which are consistent with the relevant observations, and evolve them
for ten Gyr into the future.

\begin{table*}
  %setlength{\tabcolsep}{0.55em}
  \begin{minipage}{112mm}
    \begin{center}
      \begin{tabular}{@{}lll lll lll@{}}
        Model &
        Name &
        \multicolumn{1}{c}{$\gamma_0$} &
        \multicolumn{1}{c}{$\gamma_{\infty}$}&
        \multicolumn{1}{c}{$\eta$} &
        \multicolumn{1}{c}{$M_{\infty}$}&
        \multicolumn{1}{c}{$r_{\mathrm{s}}$}&
        \multicolumn{1}{c}{$\gamma_{100\mathrm{pc}}$} &
        $M(1.8\mathrm{kpc})$
        \\
        \hline
        % name inner outer eta rho0 mass rs chisq comments
        LC & large core &
        0.07 & 4.65 & 3.7  & $8.00\times10^8$ & 1.4  & 0.1 & $4.12\times10^8$\\
        WC & weak cusp &
        0.08 & 4.65 & 2.77 & $1.23\times10^8$ & 0.62 & 0.1 & $1.03\times10^8$\\
        IC & intermediate cusp &
        0.13 & 4.24 & 1.37 & $1.51\times10^8$ & 0.55 & 0.5 & $1.03\times10^8$\\
        SC & steep cusp &
        0.52 & 4.27 & 0.93 & $1.98\times10^8$ & 0.80 & 1.0 & $1.07\times10^8$\\
      \end{tabular}
    \end{center}
    \vspace*{-1ex}
    \caption{Parameters for halo mass models (equation~\ref{eq:rho:pure}) used
      in the simulations ($\gamma_0$, $\gamma_\infty$, $\eta$, $M_\infty$,
      $r_{\mathrm{s}}$), as well as the resulting logarithmic density slope
      $\gamma(r)\equiv-\mathrm{d}\ln\rho/\mathrm{d}\ln r$ at 100\,pc and the
      mass within 1.8kpc, our lower limit for the tidal radius of
      Fornax. Radii are given in kpc and masses in \Msun.
      \label{tab:mass:models} }
  \end{minipage}
\end{table*}
%%
%%%%%%%%%%%%%%%%%%%%%%%%%%%%%%%%%%%%%%%%%%%%%%%%%%%%%%%%%%%%%%%%%%%%%%%%%%%%%%%
\subsection{Mass models for Fornax}
\label{sec:model:fornax}
\cite{WilkinsonEtAl2002} and \cite{KleynaEtAl2002} demonstrated that the
mass-anisotropy degeneracy inherent in kinematic modelling can be broken using
distribution-function modelling of sufficiently large kinematic data sets. To
take full advantage of the recent data set of more than two thousand
individual stellar velocities in Fornax~\citep{WalkerEtAl2009}, Wilkinson et
al. (in preparation) apply the Markov-Chain-Monte-Carlo (MCMC) technique to
dynamical and mass models of Fornax. The mass profile and stellar-luminosity
profile are modelled independently using spherical double-power-law profiles
of the form
\begin{equation}\label{eq:rho:pure}
  \rho_{\mathrm{model}}(r) = \rho_0 \, 
  \left(\frac{r}{r_{\mathrm{s}}}\right)^{-\gamma_0}
  \left(1+\left(\frac{r}{r_{\mathrm{s}}}
  \right)^\eta\right)^{\frac{\gamma_0-\gamma_\infty}{\eta}}.
\end{equation}
The stellar distribution functions are calculated numerically following the
approach of \cite{Gerhard1991} and \cite{Saha1992} allowing various velocity
anisotropy profiles. As in the earlier work \citep{WilkinsonEtAl2002,
  KleynaEtAl2002}, the models are compared to the data on a star-by-star
basis.

Further details of the modelling and its results will be presented
elsewhere. Here, we simply use the above MCMC model ensemble to inform our
choice of halo models. Rather than considering a single best-fit model of
Fornax, we select four mass models, each of the form~(\ref{eq:rho:pure}) but
truncated at very large radii via
\begin{equation}\label{eq:rho}
  \rho(r) = \rho_{\mathrm{model}}(r)\;
  \mathrm{sech}(r/10\mathrm{kpc}),
\end{equation}
and with parameter values as detailed in Table~\ref{tab:mass:models}. These
models span the range of models consistent with the kinematic data. The three
models WC (weak cusp), IC (intermediate cusp) and SC (steep cusp) have
parameters $\gamma_0$, $\gamma_\infty$, $\eta$, $r_{\mathrm{s}}$, and
$M_\infty$ directly taken from the MCMC chain outputs, and refer to the
highest likelihood models with density slope
\begin{equation}\label{eq:gamma}
  \gamma(r) \equiv - \frac{\mathrm{d}\ln\rho}{\mathrm{d}\ln r}
\end{equation}
of, respectively, 0.1, 0.5, and 1.0 at $r=100\,$pc.

The fourth model, LC (large core), was motivated by the recent work of
\cite{WalkerPenarrubia2011}. These authors applied a non-parametric
statistical modelling technique to two chemically distinct stellar populations
within Fornax to define the enclosed mass at the half light radii of the two
populations. The resulting model possesses a large core with near-constant
density and $\gamma\approx0.1$ for $r\lesssim500\,$pc.

\begin{figure}
  \centerline{
    \resizebox{78mm}{!}{\includegraphics{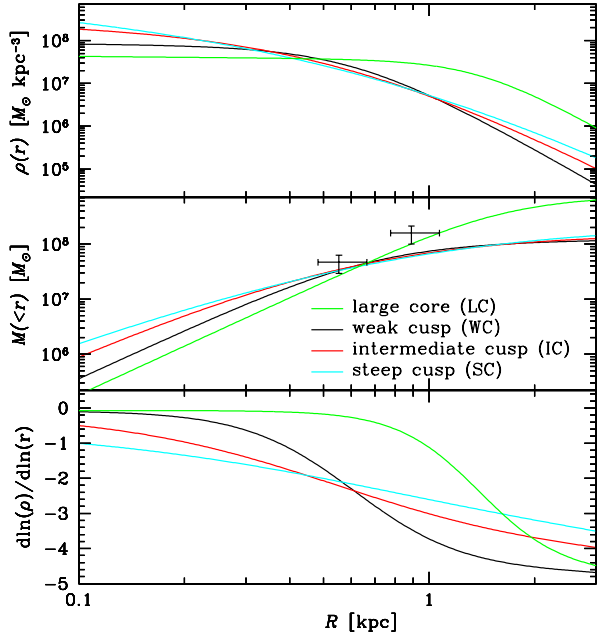}}
  }
  \caption{
    \label{fig:halomass}
    Radial profiles of density (\emph{top}), enclosed mass (\emph{middle}),
    and logarithmic density slope (\emph{bottom}) for the four mass models
    used in our simulations (see also Table~\ref{tab:mass:models}).The data
    points the middle panel correspond to the mass estimates by
    Walker \& Pe\~narrubia (2011)%\cite{WalkerPenarrubia2011}
    for two chemically distinct sub-populations.}
\end{figure}

The radial profiles of density, enclosed mass, and logarithmic density slope
of the four mass models are shown in Fig.~\ref{fig:halomass} for comparison.
These models cover a wide range of inner density slopes, including shallow
profiles, such as suggested by \citep{GilmoreEtAl2007} based on observations
of dSph galaxies, but also a steep cusp, such as predicted by cosmological
simulations \citep{DubinskiCarlberg1991, NavarroFrenkWhite1996}. Note,
however, that our models represent the overall mass distribution of Fornax
including both the stars and the dark matter.
\begin{figure}
  \centerline{
    \resizebox{78mm}{!}{\includegraphics{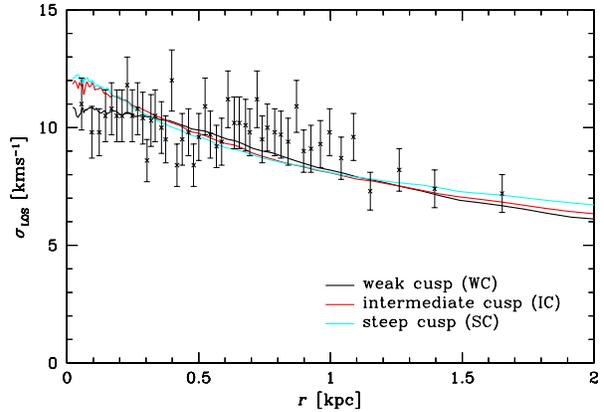}}
  }
  \caption{
    \label{fig:stellarvdisp}
    The observed stellar velocity dispersion for Fornax as a function of
    projected radius 
    (Walker et al. 2007),%\citep{WalkerEtAl2007}, 
    and the simple predictions
    obtained for the four mass models used in this study under the assumption
    of isotropic velocities. The purpose of this comparison is not to assess
    the relative merit of the mass models, but merely to demonstrate that
    their normalisations are reasonable.  }
\end{figure}

Figure \ref{fig:stellarvdisp} shows the stellar velocity dispersion for of our
three MCMC-based Fornax models (\emph{curves}) plotted together with the
observed velocity dispersion as measured by \cite{WalkerEtAl2007}. The model
velocity dispersions were calculated assuming an ergodic distribution function
with a \cite{plummer1911} density profile with core radius 668\,pc for the
stellar component, but the respective halo model for the underlying mass
distribution. In view of the fact that these simple ergodic models have not
been fitted to the velocity-dispersion data (apart from the assumed mass
models), they provide surprisingly good a description of these data. (This and
the similarity between the model predictions is exactly the reason why
inferring the mass profile from data like th\corr{ese} is hardly possible.)

The large core model (LC) has been normalised to roughly agree with the
estimates for the enclosed mass derived from two different tracer populations
by \cite{WalkerPenarrubia2011} and plotted in Fig.~\ref{fig:halomass}.

%%%%%%%%%%%%%%%%%%%%%%%%%%%%%%%%%%%%%%%%%%%%%%%%%%%%%%%%%%%%%%%%%%%%%%%%%%%%%%%
\subsection{Modelling the globular cluster system}
\label{sec:model:cluster}
If we compare the distance to the Fornax dSph with the distances to the
individual clusters in Table~\ref{tab:GCdata}, it can be seen that the
measurements of the distances are not accurate enough to provide reliable
three-dimensional locations within the dSph. We therefore draw for each
simulation random line-of-sight distance offsets to Fornax from a uniform
distribution between 0 and 2\,kpc, the approximate tidal radius of the system
\citep{WalkerPenarrubia2011}. For the velocities, we choose a similar
statistical approach by sampling the full space velocity from a bi-variate
Gaussian distribution, specified by the total velocity dispersion $\sigma$ and
the anisotropy parameter
\begin{equation}\label{eq:beta}
  \beta\equiv 1-\frac{\sigma_{\theta}^2+\sigma_{\phi}^2}{2\sigma_r^2}.
\end{equation}
This is done in such a way that for clusters GC2-5 the line-of-sight velocity
matches the observed value (which may be considered a prior for our sampling).

As the spatial distribution of the clusters is consistent with that of the
stars in Fornax, it seems reasonable to base our kinematical parameters for
the clusters on the observed stellar kinematics. The measured stellar velocity
dispersion is approximately flat over the range of radii observed
\citep{WalkerEtAl2007,Lokas2009}, and we use $\sigma=10.5\,$km\,s$^{-1}$. For
the velocity anisotropy we assume $\beta=-0.33$ as suggested for the stars
\citep{Lokas2009}, i.e.\ a mild tangential bias, which gives
$\sigma_r\approx9.5\,$km\,s$^{-1}$ and $\sigma_t\approx11\,$km\,s$^{-1}$.

With this modelling approach, in particular the wide range of initial radii
sampled, we generate many different orbits for the globular clusters, covering
the complete range of all possible orbits for them. By allowing such a wide
distribution of cluster orbits, we can explore the effects of possible narrow
choices for the cluster distribution function, such as preferably circular
orbits, afterwards by restricting our analysis correspondingly.

%%%%%%%%%%%%%%%%%%%%%%%%%%%%%%%%%%%%%%%%%%%%%%%%%%%%%%%%%%%%%%%%%%%%%%%%%%%%%%%%
\subsection{\boldmath $N$-body simulations}
\label{sec:technicalities}
For each Fornax mass model, we ran 700 $N$-body simulations, each with
different initial conditions for the five clusters.

The individual cluster positions and velocities are drawn as described in the
previous sub-section and their masses are taken from Table~\ref{tab:GCdata}.
The clusters are represented by individual massive but softened particles with
density profiles
\begin{equation} \label{eq:soft:rho}
  \rho(r) \propto(r^2+\epsilon^2)^{-7/2},
\end{equation}
where $\epsilon=5\,$pc, comparable to the cluster core radii.

To generate the $N$-body initial conditions for the Fornax mass models, we
sample positions from the density~(\ref{eq:rho}) and velocities from
self-consistent ergodic distribution functions $f(E)$, which only depend on
the specific orbital energy $E$, thus giving everywhere isotropic velocity
distributions. The forces between particles representing Fornax are softened
with softening length $\epsilon=10\,$pc.

\begin{figure*}
  \resizebox{88mm}{!}{\includegraphics{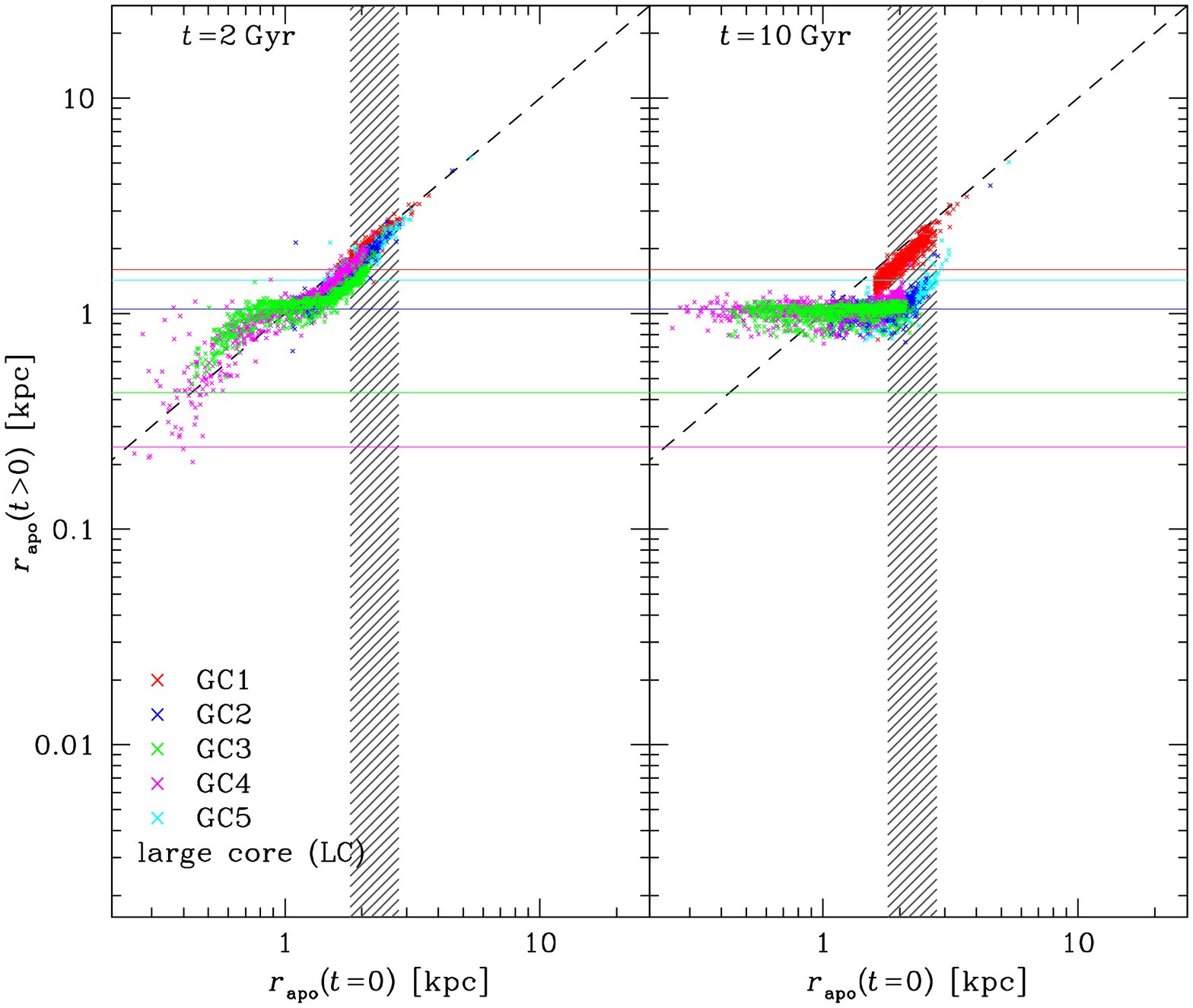}}\hfill
  \resizebox{88mm}{!}{\includegraphics{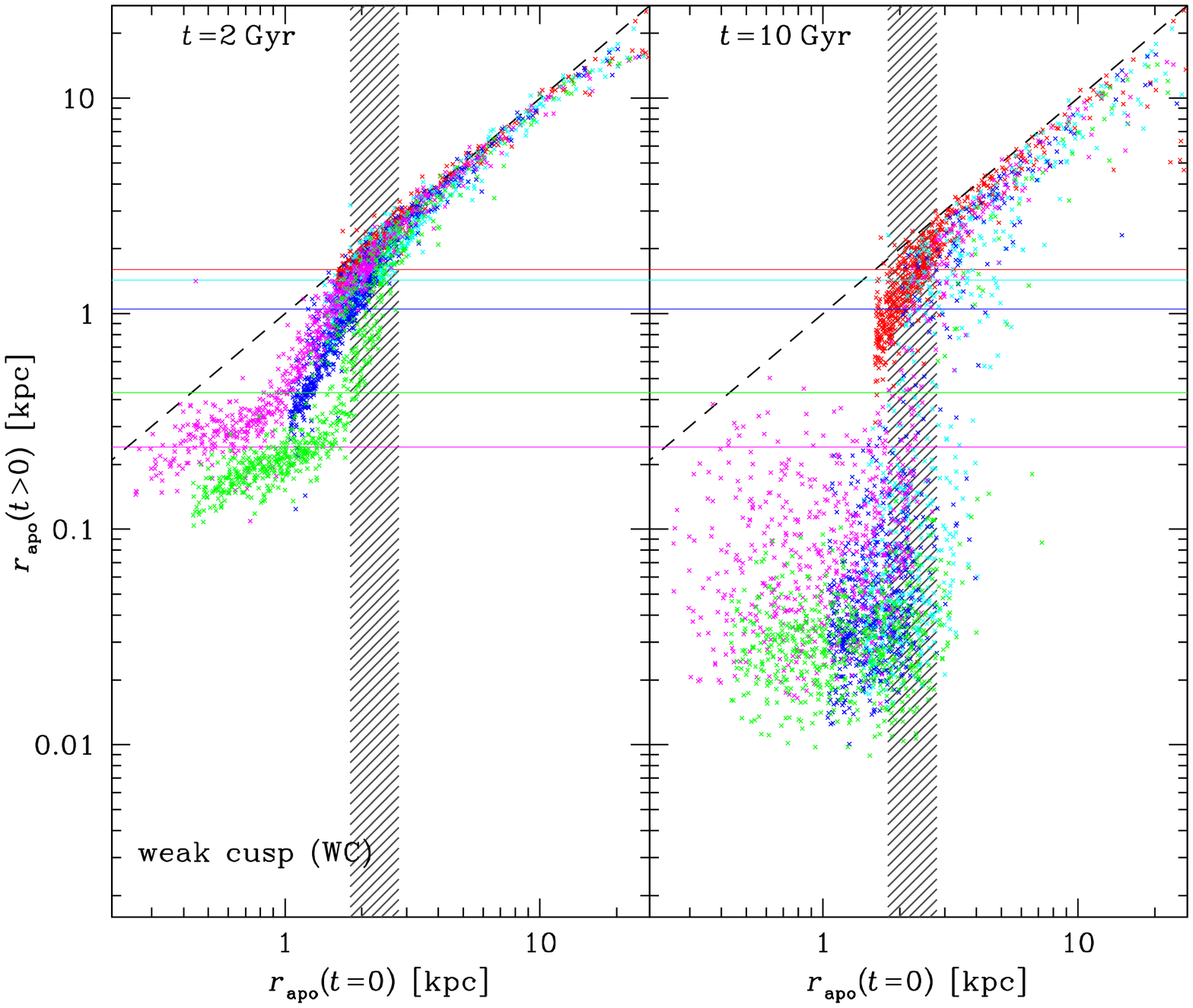}}\\[1ex]
  \resizebox{88mm}{!}{\includegraphics{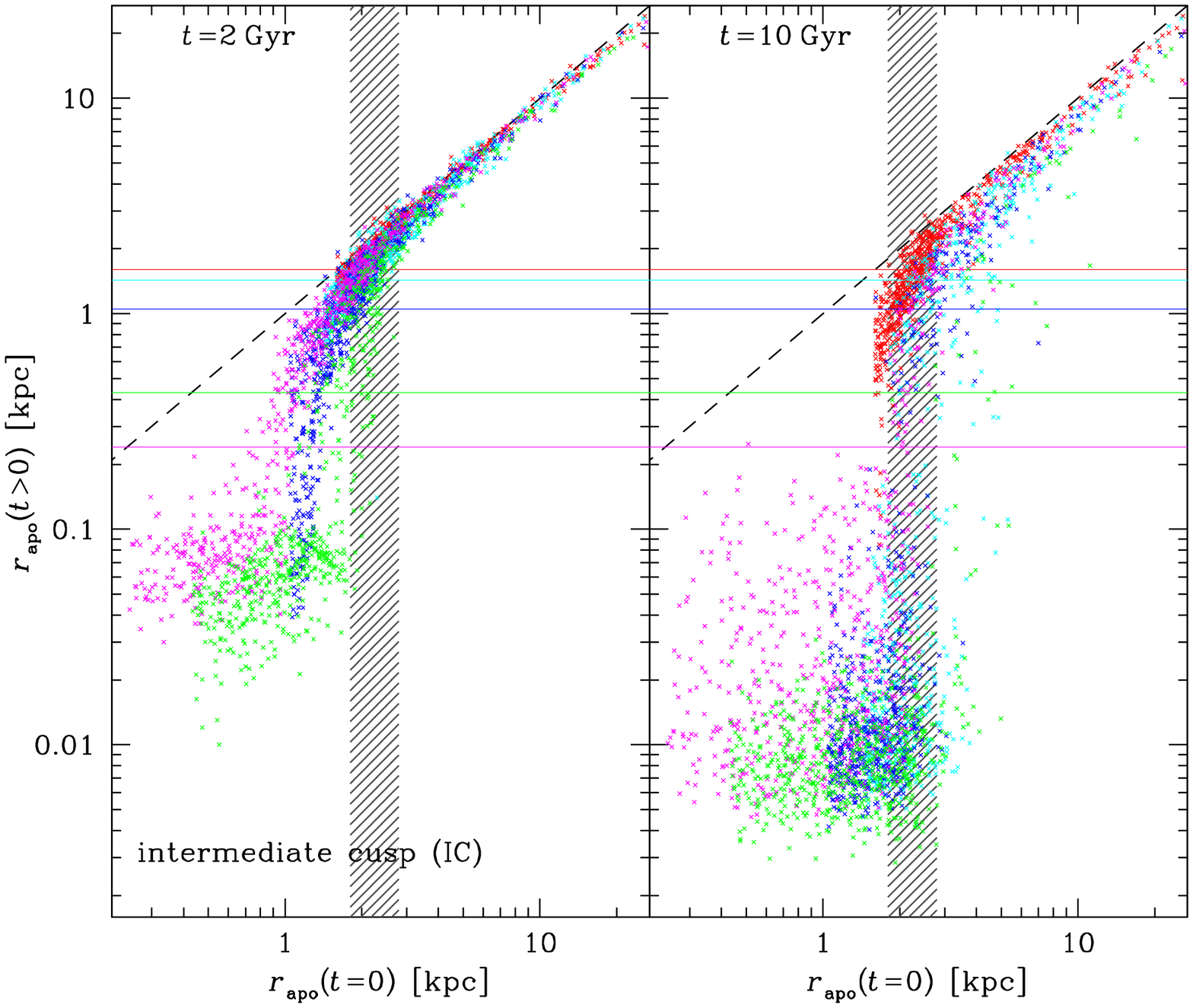}}\hfill
  \resizebox{88mm}{!}{\includegraphics{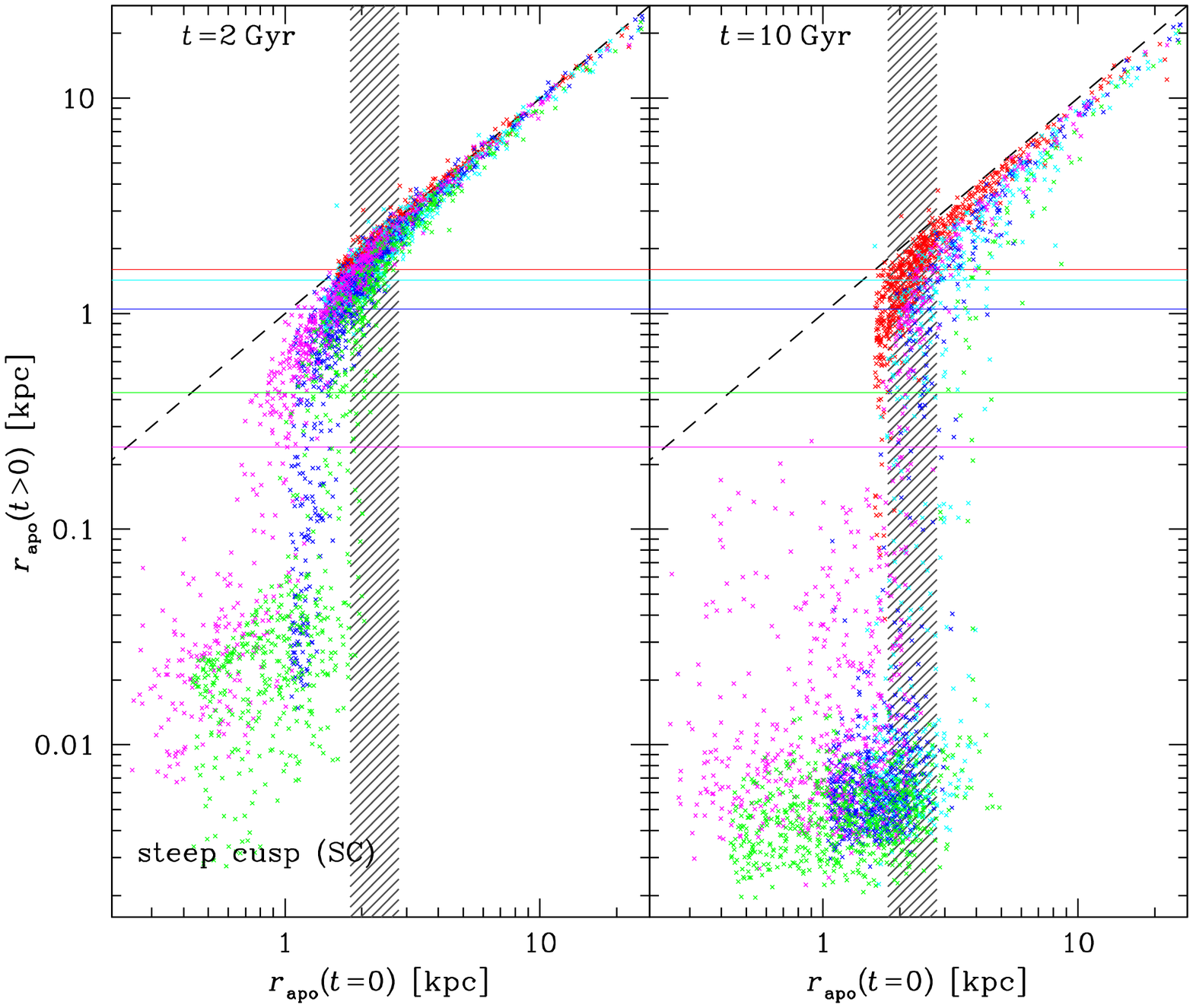}}
  \caption{Apo-centric radii $\rapo$ of the instantaneous cluster orbits after
    $t=2$ and 10\,Gyr for all simulations per halo model (as indicated)
    plotted versus $\rapo$ of the initial orbit. The shaded region indicates
    the most likely value for the current tidal radius of Fornax. Any initial
    $\rapo$ greater than that would have been strongly affected (likely
    removed from Fornax) by the Galactic tidal field (not modelled in our
    simulations). The thin horizontal lines indicate the observed projected
    radius $R_{\mathrm{obs}}$ for each cluster, which is necessarily less than
    its initial $\rapo$. Any final $\rapo$ below these lines would be
    inconsistent with the present cluster location.\label{fig:R:Ra} }
\end{figure*}
We enhance the resolution of the $N$-body model in the inner parts (where
dynamical friction occurs) by increasing the sampling probability by a factor
$g(E)^{-1}$ which is compensated by setting particle masses $\mu_i$
proportional to $g(E_i)$. We used
\begin{equation}
  g(E) \propto \frac{1+q\,r_{\mathrm{circ}}^{\eta}(E)}
  {r_{\mathrm{circ}}^{\eta}(E)+r_s^{\eta}}
\end{equation}
with $q=4$ the ratio between maximum and minimum particle mass and
$r_{\mathrm{circ}}(E)$ the radius of the circular orbit with specific energy
$E$. Testing this method for our particular purposes, we found that it allows
a reduction of $N$ to half at the same central resolution without any adverse
effects. Based on convergence tests of decaying cluster orbits (see Appendix
\ref{app:converge}), we use $N = 2\times10^6$ particles to sample each of the
Fornax mass models.

The mass ratio of the lightest background particle in all our models to the
lightest cluster is 1:1680 (in model WC) and that of the heaviest particle to
the lightest cluster 1:62 (in model LC). The clusters orbit mainly within the
high-resolution region of our models defined by the volume where the
resolution is better than would have been achieved with the same number of
particles of constant mass (within approximately 2\,kpc).

The simulations are performed using the publicly available $N$-body code
\textsf{gyrfalcON}, which uses Dehnen's (\citeyear{Dehnen2000:falcON},
\citeyear{Dehnen2002}) $\mathcal{O}(N)$ algorithm for force approximation. The
total energy conservation was typically a few parts in $10^4$.

%%%%%%%%%%%%%%%%%%%%%%%%%%%%%%%%%%%%%%%%%%%%%%%%%%%%%%%%%%%%%%%%%%%%%%%%%%%%%%%%
\section{Raw simulation results}
\label{sec:results}
As detailed at the end of \S\ref{sec:model:cluster}, our simulations cover a
wide range of initial cluster orbits, some of which may not be very realistic.
In this section we ignore any implications of our sampling of the initial
cluster orbits and simply consider the individual simulations on their own
merit.

For each simulation, we need to quantify in how much each simulated cluster
has suffered from dynamical friction and has sunken into the core of the dSph.
In this respect it is clearly better to use an orbital property instead of an
instantaneous quantity, such as radius (projected or intrinsic). Therefore, we
use as our main characteristic the apo-centric radius $\rapo$ of the
instantaneous cluster orbit. This is obtained from the cluster's instantaneous
angular momentum $L$ and specific energy $E$ as the larger of the two radii
for which
\begin{equation} \label{eq:instap}
  E = \frac{L^2}{2 r} + \Phi(r)
\end{equation}
(the smaller is the peri-centric radius). $\Phi(r)$ is the instantaneous
gravitational potential of the $N$-body model, estimated using spherical
averaging. For a small fraction of simulated clusters, the initial orbits were
unbound. Such simulations are discarded only for analysis of the affected
cluster orbit, but not for the others.

For each of the four mass models of Table~\ref{tab:mass:models},
Fig.~\ref{fig:R:Ra} plots the distributions of each cluster orbit in initial
and final $\rapo$ at $t=2\,$Gyr and $t=10\,$Gyr in the left and right
sub-panels, respectively.

%%%%%%%%%%%%%%%%%%%%%%%%%%%%%%%%%%%%%%%%%%%%%%%%%%%%%%%%%%%%%%%%%%%%%%%%%%%%%%%%
\subsection{Orbital decay after 2\,Gyr}
\label{sec:result:raw:2Gyr}
After 2\,Gyr (left sub-panels in Fig.~\ref{fig:R:Ra}) $\rapo$ is significantly
less than initially for most orbits with initial $\rapo<2-3\,$kpc, except for
model LC, which we discuss separately in \S\ref{sec:result:raw:LC}.

However, both GC1 (red) and GC5 (cyan) rarely sink substantially, and GC1 only
ever falls into the centre of Fornax for the most cusped model SC. This is not
surprising since GC1 is not only by far the lightest of the five clusters (see
Table~\ref{tab:GCdata}), but also the one furthest away (in projection) from
the centre of Fornax and hence less likely to pass through high density
regions. As such it requires most dynamical friction to fall into Fornax, but
will suffer least, since drag force $\propto\mathrm{mass}^2\rho$. Though GC5
is five times more massive, it has the second greatest projected distance from
the centre of Fornax and hence also suffers significantly less friction than
the other clusters for most orbits sampled.

The remaining three clusters GC2-4 \corr{are all dragged inwards} when
initially $\rapo<2-3\,$kpc\corr{, presenting the Fornax timing problem}.
For all of these clusters, orbits with initial $\rapo\lesssim1\,$kpc show
similar effects of dynamical friction, presumably because orbits with
apo-centres as small as that spend sufficient time in high-density regions to
suffer substantially from dynamical friction.

Since $\Robs\le\rapo$, an initial $\rapo\sim1\,$kpc is the absolute possible
minimum for GC2, which is observed at $\Robs=1.05\,$kpc.  Consequently GC2
only rarely falls in as much as GC3 and GC4.

Apart from the initial $\rapo$, the infall of these clusters depends most
strongly on the inner density profile of the background mass distribution.
Model SC shows the greatest effect of dynamical friction on the clusters. For
GC3 and GC4, a significant proportion of the simulations find their
instantaneous apo-centres inside 30\,pc. GC2 does not show such a marked
effect but a number of simulations already have an apo-centre inside 100 pc.

By contrast, model WC shows significantly reduced dynamical friction. GC3
again is most affected. However, even in the most extreme cases, the
apo-centres have decayed to no less than $\sim100\,$pc from the centre. In
this model the logarithmic density slope $\gamma$ (equation \ref{eq:gamma})
was initially only 0.1 at 100\,pc, which implies a near-flat density profile
within this radius. As expected, model IC is intermediate between models WC
and SC.

\begin{figure}
  \centerline{
    \resizebox{78mm}{!}{\includegraphics{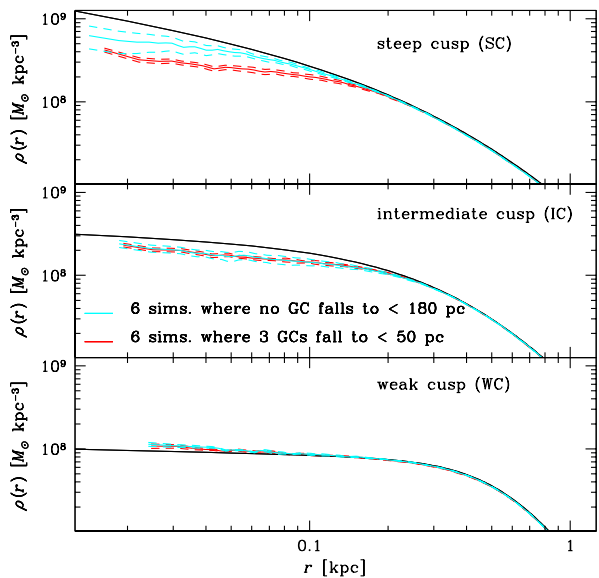}}
  }
  \caption{
    \label{fig:denscore}
    Ths initial (\emph{black}) and final ($t=10\,$Gyr, \emph{coloured})
    density profile of six halo models (stacked $N$-body models) from
    simulations where either three clusters sunk into Fornax or
    none. The \emph{dashed} curves represent $\pm1\sigma$.
  }
\end{figure}
%%
%%%%%%%%%%%%%%%%%%%%%%%%%%%%%%%%%%%%%%%%%%%%%%%%%%%%%%%%%%%%%%%%%%%%%%%%%%%%%%%%
\subsection{Orbital decay after 10\,Gyr}
\label{sec:result:raw:10Gyr}
After 10\,Gyr, the trends shown at 2\,Gyr are amplified. Four of the globular
clusters fall into Fornax for most simulations. Only GC1 still shows not much
evidence of migrating to the centre of Fornax. All cluster have a small
fraction of simulations where they remain at large $\rapo$. This requires the
initial $\rapo$ to be large as well. In general, $\rapo$ decreases at least
slightly, even when initially large (the few significant increases of $\rapo$
are caused by cluster encounters).

For the most cusped model SC and to some degree for model IC too, GC3, GC4 and
even GC2 can obtain values for $\rapo$ down to close to their softening length
of 5\,pc---i.e.\ they are essentially at the very centre of the galaxy. For
model WC, this is not the case, i.e.\ none of the clusters can reach the very
centre of the galaxy in this case. However, the vast majority of simulations
do not obtain such very small $\rapo$.

We observe that the distributions of $\rapo$ after 10\,Gyr are quite similar
between models WC, IC, and SC (in particular the latter two), when GC3 has
settled at $\rapo<100\,$pc for most of our simulations, in fact all
simulations with initial $\rapo<2.8\,$kpc, and the distributions for $\rapo$
of GC4 are remarkably similar. For the most massive GC3, this similarity
reflects the fact that the cluster has essentially sunken to the very centre
of the galaxy.

However, this cannot explain the similarities for GC4, which we think is
caused by a similarity of the background mass profiles after 10\,Gyr. The
action of cluster dynamical friction causes dynamical heating of the
background particles dragging the clusters. This in turn erases the initial
central density cusp \citep{ReadEtAl2006a, ColeDehnenWilkinson2011}.
\cite{GoerdtEtAl2010} discuss the formation of a density core in this way and
provide an empirical formula for the expected core size created as the
clusters fall to the centre. For model WC this predicts a core radius of
262\,pc which is larger than our stalling radii but agrees within a factor of
a few. However this formula fails to predict the behaviour of the clusters in
the IC and SC models as it gives stalling radii of 176 pc and 100 pc
respectively.

In Fig.~\ref{fig:denscore}, we plot the density profiles of the final models
for cases with and without cluster infall. For the steep and intermediate cusp
models, the central density profiles are significantly reduced and in fact
have become much more similar to each other. However, only for model SC is
this reduction slightly stronger when clusters have reached the core of
Fornax. In all cases, the density still keeps on increasing inwards, and hence
does not really correspond to a constant-density core in the classical
sense\footnote{For model WC, the density actually \emph{increases} slightly at
  $r<100\,$pc. This is presumably caused by a slight instability of this
  model, which has $\diff f/\diff E>0$ for some range of specific energies $E$
  and hence may be unstable \citep[see][\S5.5]{BinneyTremaine2008}.}.

\begin{figure}
  \centerline{
    \resizebox{78mm}{!}{\includegraphics{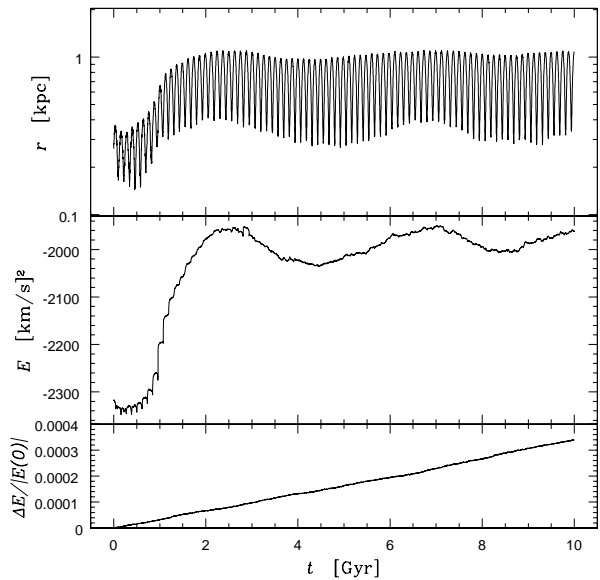}}
  }
  \caption{
    \label{fig:buoyGC4}
    The evolution of the orbit of GC4 in one of our simulations with the large
    core (LC) model. The top and middle panels show the development of the
    orbital radius and specific energy for GC4, respectively. The latter grows
    after approximately 2\.Gyr as the orbit moves outwards. The overall
    conservation of energy (bottom panel) is not affected by this orbital
    change.}
\end{figure}
%%
%%%%%%%%%%%%%%%%%%%%%%%%%%%%%%%%%%%%%%%%%%%%%%%%%%%%%%%%%%%%%%%%%%%%%%%%%%%%%%%%
\subsection{The large-core model}
\label{sec:result:raw:LC}
The large core model (LC) shows very unusual behaviour. After 2\,Gyr, all
globular clusters have apo-centres closely clustered together with a strong
peak at $\sim1\,$kpc. This becomes even more pronounced at 10\,Gyr. Detailed
examination of the cluster orbits in individual simulations shows two
interesting behaviours. First, orbits with initial $\rapo\gtrsim900\,$pc decay
move in quite rapidly (mostly in less than 2\,Gyr) to $\rapo\sim900\,$pc,
where they stall. This confirms the work of \cite{GoerdtEtAl2006} and
\cite{ReadEtAl2006a} which showed that massive satellites orbiting outside of
an harmonic density core stall at the edge of the core. This behaviour is
believed to be due to the reduction of dynamical friction due to the resonant
effects of particles in the harmonic core.

Second, any cluster which has an initial orbit within the harmonic core move
\emph{out} to the edge of the core. Fig.~\ref{fig:buoyGC4} shows the evolution
of the orbit of GC4 in one of our simulations. As can be seen the cluster
orbit absorbs energy as it expands. The overall conservation of energy for the
simulation is not affected by this change and energy is conserved to
approximately 3 parts in $10^4$ during the simulation. This behaviour is
unexpected and we do not believe that it has been reported previously, though
there is some evidence for orbital radii expanding again after falling in at
the edge of harmonic cores in what has been called the `kickback effect'
\citep{GoerdtEtAl2010, Inoue2009}. We will discuss this further in section
\ref{sec:conclusions}.

%%%%%%%%%%%%%%%%%%%%%%%%%%%%%%%%%%%%%%%%%%%%%%%%%%%%%%%%%%%%%%%%%%%%%%%%%%%%%%%%
\subsection{Summary}
\label{sec:result:raw:sum}
The raw empirical results from our simulations can be summarised as follows.
\begin{enumerate}
  \item Cluster orbits with large initial $\rapo$ are not significantly
    affected by dynamical friction, because these orbits spend no or too
    little time in high-density regions, where frictional drag is exerted.
    However, most simulations which initial $\rapo$ less then the current
    tidal radius of Fornax suffer from dynamical friction.
  \item Cluster GC3 is most likely affected by dynamical friction, followed by
    GC4 and GC2, while clusters GC1 and GC5 are least likely affected after
    2\,Gyr. This ordering is expected from the masses and initial projected
    radii of the clusters.
  \item For all except model LC (large core), cluster GC3 always reaches the
    core of Fornax within 10\,Gyr (unless its initial orbit was
    unrealistically large with $\rapo>2.8\,$kpc beyond the present-day tidal
    radius of Fornax)\corr{, constituting the long-term timing problem}.
  \item The effect of dynamical friction at 2\,Gyr is increasing with the
    central mass density from model WC to SC, as expected from Chandrasekhar's
    dynamical friction formula.
  \item The effect of dynamical friction after 10\,Gyr is more similar for the
    three halo models with weak to steep cusps than after 2\,Gyr. This
    similarity can be understood, at least qualitatively, by stalling of
    dynamical frictions as consequence of core formation.
  \item Model LC shows no effect of dynamical friction, but rather the
    opposite: `dynamical buoyancy', when the clusters are pushed out of the
    core. Like dynamical friction, this effect appears strongest for the most
    massive cluster.
\end{enumerate}
In particular, for a CDM-type steep cusp (model SC), almost all simulations
with initial $\rapo<2.8\,$kpc (Fornax's tidal radius) suffer significantly
from dynamical friction within 10\,Gyr, or even within only 2\,Gyr. This
reflects the Fornax timing problem and is in contradiction with the claims of
\cite{AngusDiaferio2009} (as discussed in the introduction), possibly because
our sampling discourages very nearly circular orbits, but the usage of
Chandrasekhar's simple formula by these authors certainly plays a role too.

%%%%%%%%%%%%%%%%%%%%%%%%%%%%%%%%%%%%%%%%%%%%%%%%%%%%%%%%%%%%%%%%%%%%%%%%%%%%%%%%
%
\begin{figure*}
  \resizebox{88mm}{!}{\includegraphics{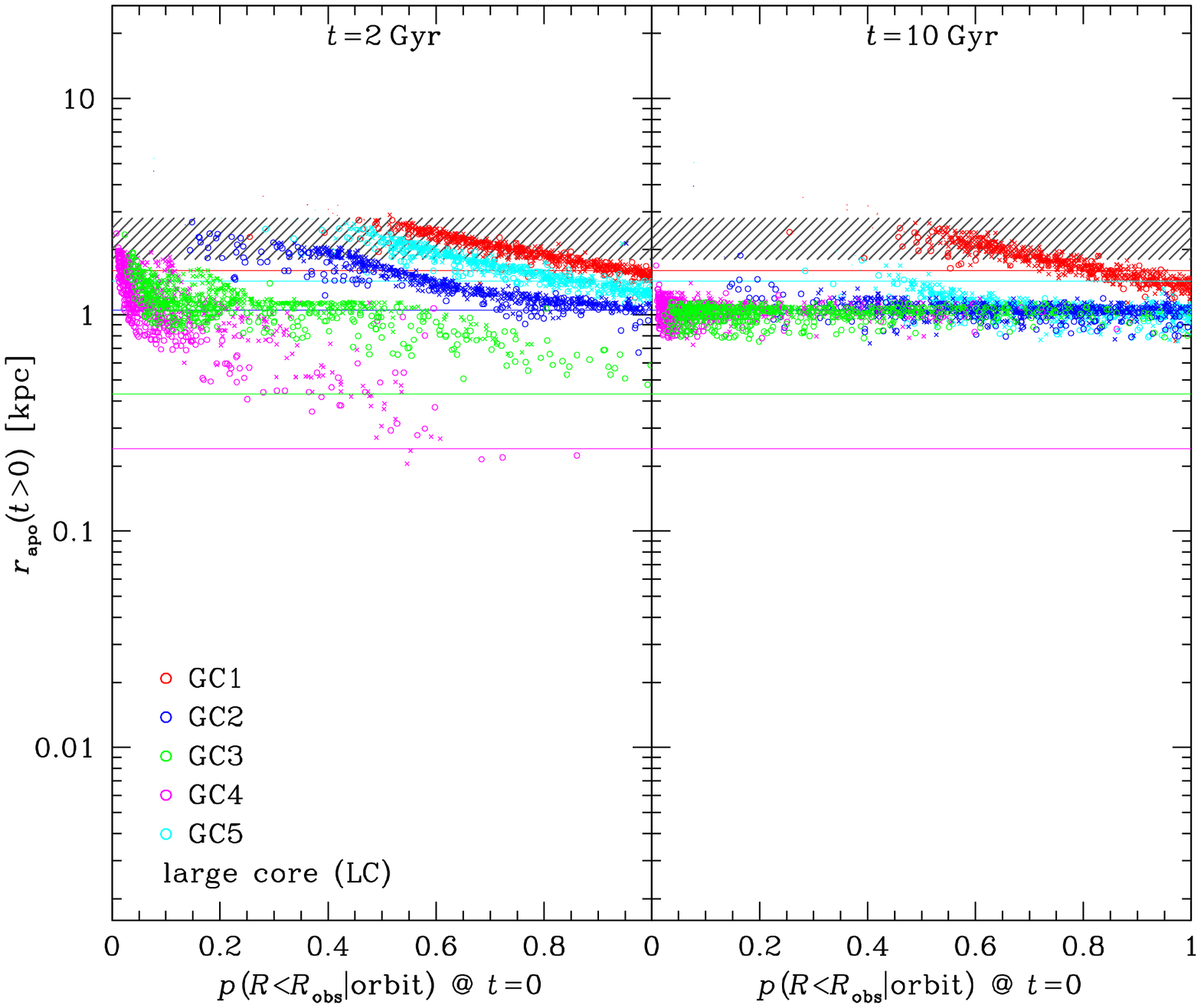}}\hfill
  \resizebox{88mm}{!}{\includegraphics{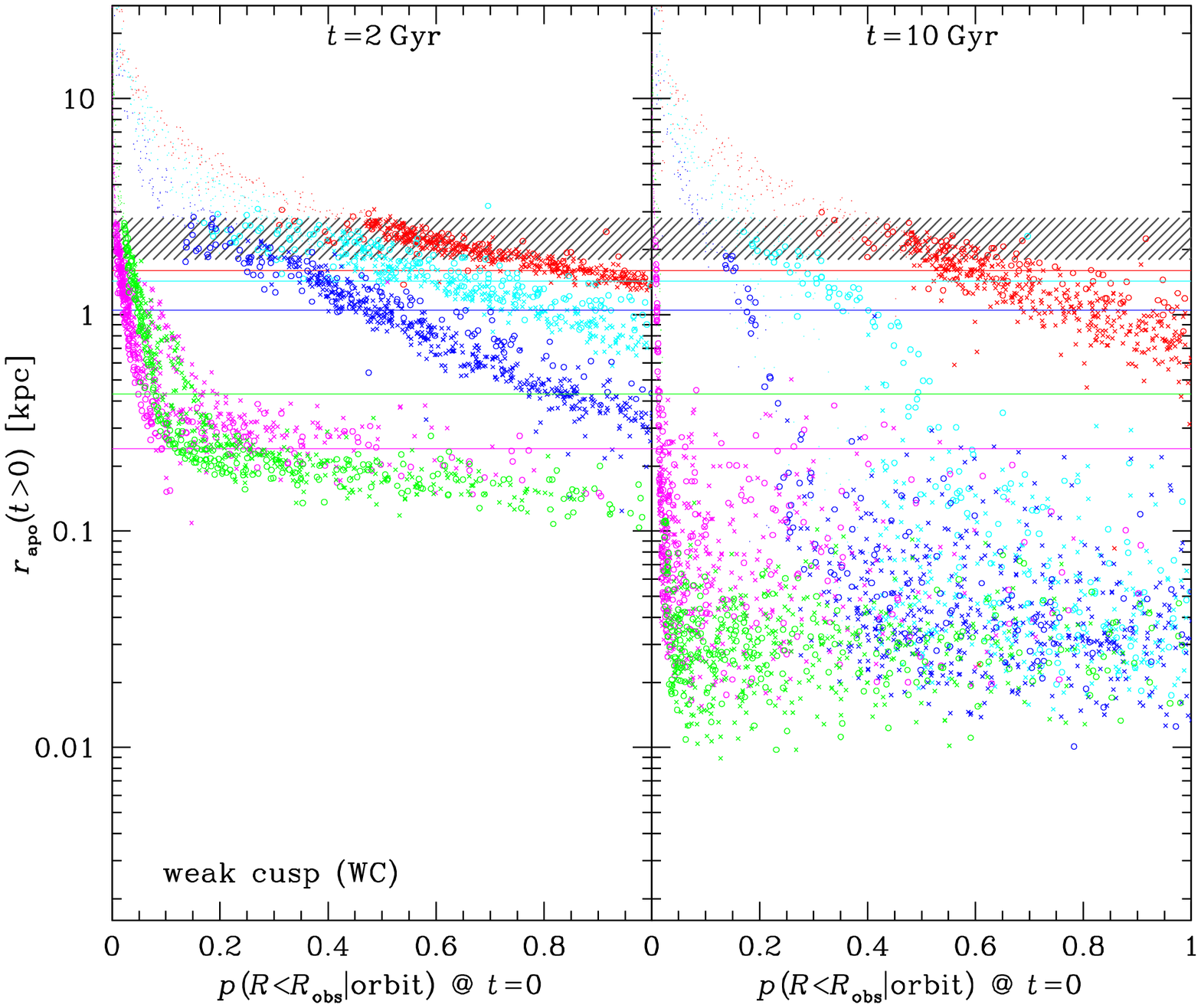}}\\[1ex]
  \resizebox{88mm}{!}{\includegraphics{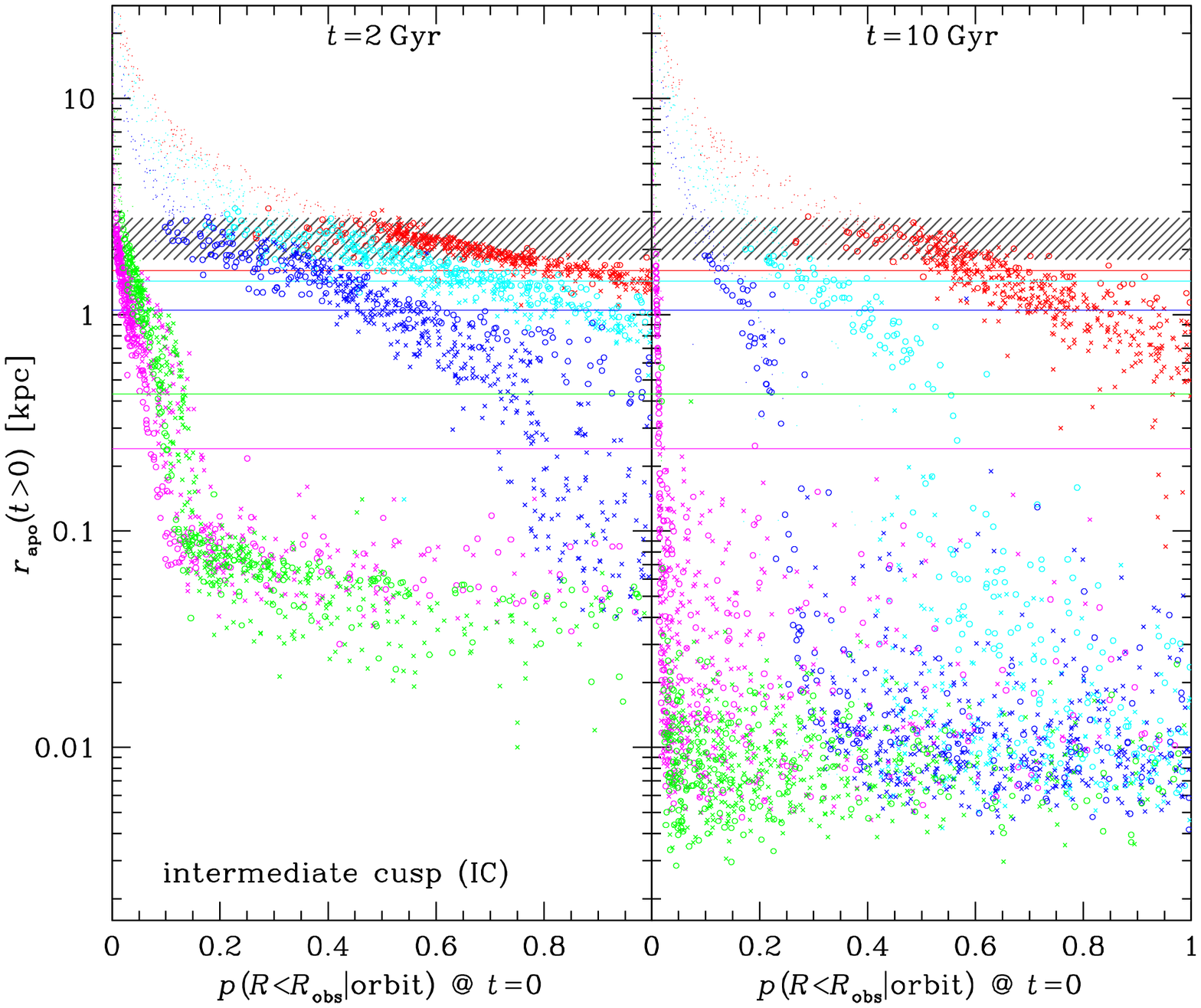}}\hfill
  \resizebox{88mm}{!}{\includegraphics{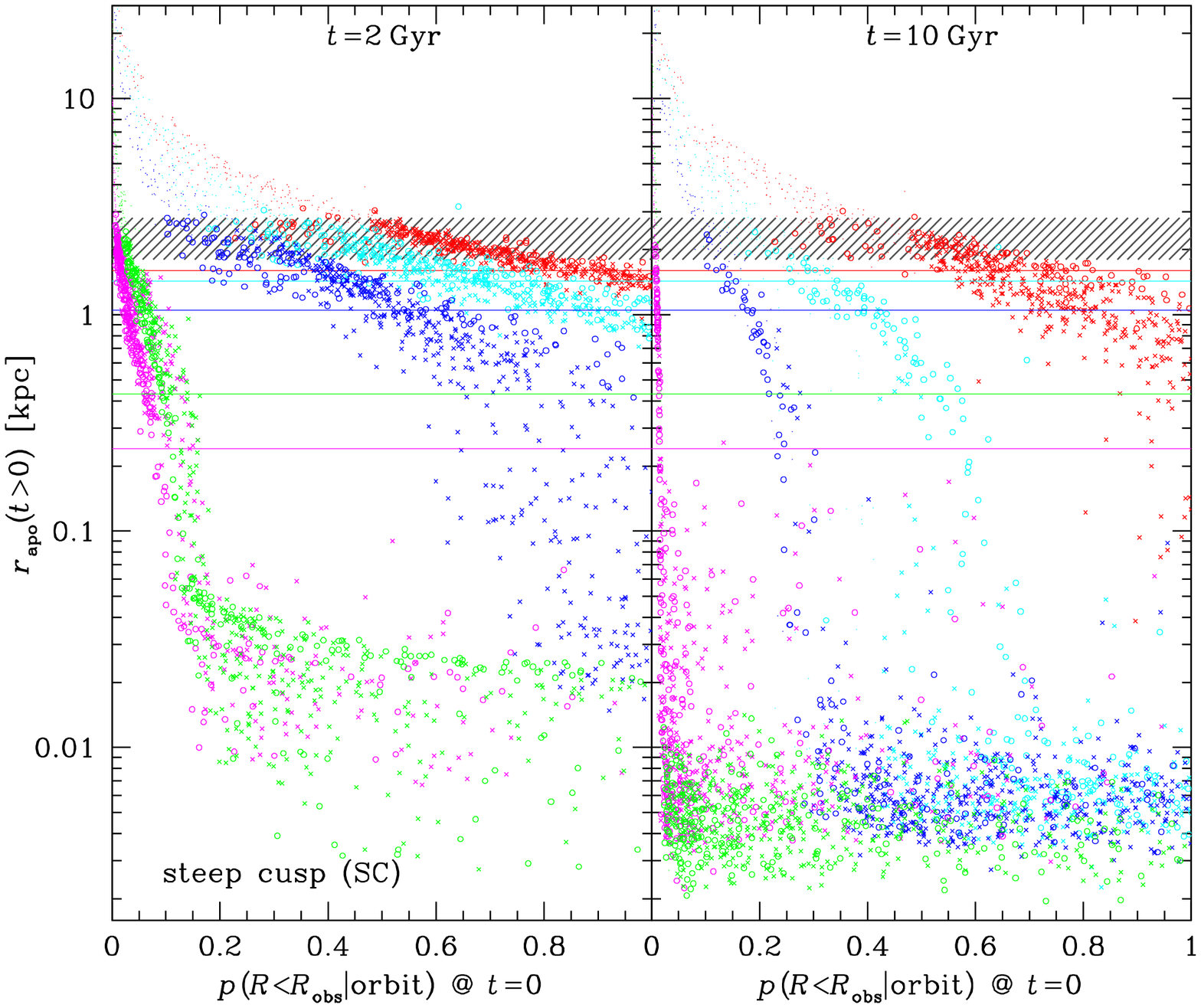}}
  \caption{Instantaneous apo-centric radius $\rapo$ at $t=2$ and 10\,Gyr (as
    in Fig.~\ref{fig:R:Ra}) for all simulated clusters in a given halo model
    (as indicated) plotted versus the fraction $p(R\le\Robs|\mathrm{orbit})$
    of orbital phases and projections for which $R\le\Robs$ on the initial
    cluster orbit. A uniform distribution of
    $p(R\le\Robs|\mathrm{orbit})\in[0,1]$ corresponds to an unbiased
    distribution of orbital phases and projection angles.  \emph{Dots} are for
    orbits with initial $\rapo>2.8\,$kpc, \emph{crosses} for eccentricity
    $e>0.4$ and \emph{circles} for $e<0.4$.  The horizontal lines indicate
    $\Robs$ and the shaded band indicates the estimates for Fornax's tidal
    radius.
    \label{fig:p:Ra}
  }
\end{figure*}
\section{The probability of cluster sinking}
\label{sec:sinking}
Our results presented in the previous Section and Fig.~\ref{fig:R:Ra}
demonstrate that the more massive clusters GC3 and GC4 will fall into the core
of Fornax within 2\,Gyr or less, unless either the mass distribution of Fornax
has a core or the cluster has an initial orbit with large apo-centric radius
$\rapo$. This latter scenario requires that the observed cluster position is
either near peri-centre (special orbital phase) or has intrinsic radius much
larger than projected (special projection geometry). \corr{Either is}
atypical, implying that this scenario is inherently unlikely.  In this
\corr{section} we try to quantify just how unlikely.

Since we know neither the current orbital phase nor projection geometry for
any of the globular clusters, our most sensible \corr{approach is to
  marginalise over both, assuming uniform distributions.} However, as we will
see, our sampling of initial cluster position and velocity did not generate a
uniform sampling in these quantities. As a consequence, any quantitative
statements, based on the raw results, about the probability of cluster infall
will be biased. In order to eliminate this bias, we must emulate a uniform
sampling of initial orbital phases and projection geometries.

To this end, we need a quantity for each simulated cluster which would follow
a known distribution were orbital phase and projection angle drawn randomly. A
natural such quantity is the fraction
\[
p(R\le\Robs|\mathrm{orbit})
\]
of orbital phases and projection geometries for which the projected radius $R$
is smaller than actually observed for the corresponding initial cluster orbit
(see Appendix~\ref{app:prob} for a formula and its derivation). Under our
basic assumption of random orbital phase and projection,
$p(R\le\Robs|\mathrm{orbit})$ is uniformly distributed between 0 and 1, in
particular $p(R\le0|\mathrm{orbit})=0$ and $p(R\le\rapo|\mathrm{orbit})=1$. We
can therefore emulate a uniform distribution in orbital phase and projection
geometry by a uniform distribution in
$p(R\le\Robs|\mathrm{orbit})$\footnote{In addition to
  $p(R\le\Robs|\mathrm{orbit})$ one may also utilise the fraction
  $p(\upsilon_z\le\upsilon_{\mathrm{los}}|\Robs,\mathrm{orbit})$ of the
  line-of-sight velocit\corr{y to be less} than observed. This fraction should
  also be uniformly distributed \emph{independently} of
  $p(R\le\Robs|\mathrm{orbit})$, and thus allow an additional independent
  constraint. However, its computation is more involved and it seems unlikely
  that much would be gained from it, mainly because of the relatively large
  uncertainty of the observed $\upsilon_{\mathrm{los}}$.}.

In Fig.~\ref{fig:p:Ra}, we plot for each simulated cluster and each halo mass
model the instantaneous apo-centric radius $\rapo$ after 2 and 10\,Gyr against
$p(R\le\Robs|\mathrm{orbit})$. For all mass models, there is a clear
correlation between $p(R\le\Robs|\mathrm{orbit})$ and $\rapo$ at later times
in the sense that larger $\rapo(t>0)$ are achieved only when the initial
projected radius was relatively small for the initial orbit. This makes
perfect sense: for small $p(R\le\Robs|\mathrm{orbit})$ the orbit spends most
of its time at $r>\Robs$ with little dynamical friction.

What is somewhat surprising, however, is how strong the correlation between
$p(R\le\Robs|\mathrm{orbit})$ and $\rapo(t>0)$ actually is, given that the
initial orbits cover a wide range of eccentricities. For simulated clusters
with initial $\rapo<2.8\,$kpc, we have split their initial orbits into low and
high eccentricity
\begin{equation} \label{eq:ecc}
  e = \frac{\rapo-\rperi}{\rapo+\rperi}
\end{equation}
with open symbols and crosses in Fig.~\ref{fig:p:Ra} corresponding to $e<0.4$
and $e>0.4$, respectively.  For the halo models IC and SC, we can see some
differentiation between these two groups of initial orbits, in particular at
$t=2\,$Gyr, in the sense one would expect: eccentric orbits obtain smaller
$\rapo$ (because they have smaller initial $\corr{\rperi$} and hence suffer
more dynamical friction) unless the observed $R$ was initially untypically
small (when they spend most of their time at large radii).

As is evident from Fig.~\ref{fig:p:Ra}, the distributions of
$p(R\le\Robs|\mathrm{orbit})$ from our simulations are not uniform:
there are many more simulations with small
$p(R\le\Robs|\mathrm{orbit})$, in particular for clusters with small
$\Robs$, such as GC3 and GC4\footnote{For model LC, no simulation has small
  $p(R\le\Robs|\mathrm{orbit})$ for clusters with $\Robs>1\,$kpc. This is a
  consequence of the larger mass of this model which implies that our sampling
  did not obtain any nearly unbound orbits with large initial $\rapo$.}. This
non-uniformity is simply a consequence of our sampling procedure, which
favours orbits with $\rapo\gg\Robs$ (in particular for clusters with small
$\Robs$, such as GC3 and GC4) resulting in non-uniform sampling of orbital
phase and projection geometry and hence introducing a bias.

Since $\rapo(t>0)$ is strongly correlated with $p(R\le\Robs|\mathrm{orbit})$,
we can read off Fig.~\ref{fig:p:Ra} the unbiased probabilities for infall into
Fornax. For example, in model SC $\rapo(t=2\, \mathrm{Gyr})>200\,$pc for GC4
\corr{(magenta)} in almost all simulations with
$p(R\le\Robs|\mathrm{orbit})<0.1$ but hardly for any simulation at larger
$p(R\le\Robs|\mathrm{orbit})$, implying that cluster GC4 will have
$\rapo>200\,$pc after 2Gyr with $\sim10\%$ probability in halo model SC.

A more quantitative procedure for removing the bias of our non-uniform
sampling of orbital phase and projection is to give each simulated cluster
orbit a weight such that the weighed distributions of simulated orbits have,
or at least are consistent with, a uniform sampling. Obviously, the weight to
choose is simply the \corr{reciproce} of the actual frequency
$f\big(p(R\le\Robs|\mathrm{orbit})\big)$ of fractions
$p(R\le\Robs|\mathrm{orbit})$ for all simulations of the same cluster in a
particular halo model. The probability for, say
$\rapo(t=2\,\mathrm{Gyr})>200\,$pc for GC4, is then obtained as the weighted
fraction of simulations that obtain $\rapo(t=2\,\mathrm{Gyr})>200\,$pc for
this cluster.

\begin{table}
  \setlength{\tabcolsep}{0.6em}
  \begin{center}
    \begin{tabular}{@{}ll@{\;}l@{\;}ll@{\;}l@{\;}ll@{\;}l@{\;}l@{}}
      \multicolumn{1}{c}{} &
      \multicolumn{3}{c}{any $e$} &
      \multicolumn{3}{c}{$e<0.4$} &
      \multicolumn{3}{c}{$e>0.4$}\\
      model & 
      1\,Gyr & 2\,Gyr & 10\,Gyr & 
      1\,Gyr & 2\,Gyr & 10\,Gyr & 
      1\,Gyr & 2\,Gyr & 10\,Gyr \\
      \hline
      & \multicolumn{9}{c}{
        probability for no cluster with $\rapo<100\,$pc} \\
      \hline
      WC &1     &1     &$<$0.001 &1     &1     &$<$0.001 &1     &1     &$<$0.001  \\
      IC &0.17  &0.032 &$<$0.001 &0.15  &0.022 &$<$0.001 &0.19  &0.037 &$<$0.001  \\
      SC &0.10  &0.017 &$<$0.001 &0.082 &0.014 &$<$0.001 &0.11  &0.020 &$<$0.001  \\
      \hline
      & \multicolumn{9}{c}{
        probability for no cluster with $\rapo<200\,$pc} \\
      \hline
      WC &0.92  &0.32  &$<$0.001 &0.87  &0.29  &$<$0.001 &0.97  &0.37  &$<$0.001  \\
      IC &0.087 &0.015 &$<$0.001 &0.049 &0.011 &$<$0.001 &0.11  &0.018 &$<$0.001 \\
      SC &0.076 &0.014 &$<$0.001 &0.063 &0.011 &$<$0.001 &0.086 &0.015 &$<$0.001
    \end{tabular}
  \end{center}
  \caption{\label{tab:prob:sinking} Probability, estimated as the fraction of
    orbital phases and projection geometries (consistent with the observed
    projected cluster positions) that no cluster with initial $\rapo<2.8\,$kpc
    (Fornax's tidal radius) has reached $\rapo<100\,$pc or $200\,$pc after 1,
    2, or 10\,Gyr of simulation, depending on the mass model and the
    eccentricity of the initial orbit.}
\end{table}
Combining this for all clusters, we estimate the probability that none of the
five cluster \corr{when initially on orbit} with $\rapo<2.8\,$kpc (our upper
limit for the tidal radius of Fornax) will fall into Fornax, in the sense that
$\rapo<100\,$pc or $<200\,$pc, within 1, 2, or 10\,Gyr. The results for each
halo model are given in Table~\ref{tab:prob:sinking} (except for model LC when
no cluster ever falls into Fornax). Columns 2-4 give the probabilities based
on all of our simulations. If we restrict the analysis to high or low
eccentricities (columns 5-7 and 8-10, respectively) or to orbits with initial
$\rapo<1.8\,$kpc (not given in Table~\ref{tab:prob:sinking}), the results are
hardly altered.

In other words: these probabilities depend mainly on the halo model and only
weakly on the distribution function of the cluster orbits. This is a direct
consequence of the tight relation, seen in Fig.~\ref{fig:p:Ra}, between
$p(R\le\Robs|\mathrm{orbit})$ and $\rapo$ at later times. This result implies
that we do not need to investigate further the implications of different
distributions functions for the initial cluster orbits.

%%%%%%%%%%%%%%%%%%%%%%%%%%%%%%%%%%%%%%%%%%%%%%%%%%%%%%%%%%%%%%%%%%%%%%%%%%%%%%%%
\section{Conclusions} \label{sec:conclusions}
The Fornax galaxy is unique among the Milky Way dSphs in having five surviving
globular clusters. These clusters are metal poor and very old -- comparable
with the oldest globular clusters in the Milky Way. There is a well-known
timing problem for these clusters: over a Hubble time or even a much shorter
time scale dynamical friction should drag them to the centre of Fornax, where
they would form a nuclear star cluster. Yet no such stellar nucleus is
observed in Fornax or, in fact, in any other dSph galaxy.

In this study, we have extended previous work on what the current location of
Fornax's globular clusters can tell us about Fornax's mass distribution. We
explored four different mass models for the underlying potential in Fornax; we
used the latest data for Fornax's globular clusters as constraints on their
phase space distribution; and we ran thousands of $N$-body models to sample
the uncertainties in the cluster position and velocity distribution for each
of our five mass models. This large search of the available parameter space
allowed us to hunt for viable solutions to Fornax's timing problem.

%%%%%%%%%%%%%%%%%%%%%%%%%%%%%%%%%%%%%%%%%%%%%%%%%%%%%%%%%%%%%%%%%%%%%%%%%%%%%%%%
\subsection{Caveats}
Our models all assume a spherical mass distribution for Fornax, yet the stars
are observed to be ellipsoidal in projection, implying that the true intrinsic
distribution is aspherical, most likely triaxial. The parameter space of such
configurations is considerably larger (and the freedom in cluster orbital
projections smaller), potentially allowing more solutions to the timing
problem. However, previous studies suggest that, while dynamical friction is
on average stronger on box orbits than on loop
orbits\citep*{CapuzzoDolcettaVicari2005}, triaxiality has no sifnificant
overall effect on the strength of dynamical friction
\citep*{CoraVergneMuzzio2001,Sachania2009}. This can be understood as
cancellation of two opposing effects. First, because angular momentum is not
conserved, there is no barrier for eccentric orbits to reach arbitrary small
radii and hence high densities and strong drag. Second, also because angular
momentum is not conserved, successive peri-centric passages have different
radius and density, such that an orbiting cluster only rarely suffers very
high drags.

Another minor caveat of our modelling is our ignorance of the tidal field of
the Milky-Way potential. In principle, it would be trivial to have our
$N$-body models orbit in a static model for the Galactic potential. However,
such a procedure would add even more only weakly constrained parameters
without adding significant benefits. With our existing models, we take the
Galactic tidal effects into account by discounting any simulated cluster
orbits with apo-centric radius beyond (best estimates of) Fornax's tidal
radius.

Finally, by modelling each globular cluster as a massive particle, we have
ignored their inner dynamics and tidal interaction with Fornax. Mass-loss
rates for these clusters, however, are likely to be too small to significantly
reduce their mass \citep{GoerdtEtAl2010}. Tidal disruption is arguably only
relevant for GC1, as discussed in \S\ref{sec:GC1} below.

%%%%%%%%%%%%%%%%%%%%%%%%%%%%%%%%%%%%%%%%%%%%%%%%%%%%%%%%%%%%%%%%%%%%%%%%%%%%%%%%
\subsection{Solutions to the timing problem}
Our simulations demonstrate that for normal mass models (WC, IC, or SC) for
Fornax the infall of clusters GC3 and GC4 within a Hubble time is unavoidable
and the infall of all clusters except GC1 most likely. This constitutes the
\emph{long-term} timing problem in the sense of hypothesis \ref{hyp:long} from
page~\pageref{hyp:long}, which is only really a problem if one assumes that
the present globular cluster distribution of Fornax is representative of the
distant past.

Only our large-core model (LC) avoids this problem completely. Rather than
dynamical friction, this model shows `dynamical buoyancy', when clusters are
pushed out of the core. Following \cite{TremaineWeinberg1984}, this is likely
caused by this model's inverted distribution function, when $\diff f/\diff
E>0$ over a significant range of orbital energies $E$. Such
models\footnote{For self-gravitating systems, an inverted distribution
  function typically occurs if the transition between a near-constant density
  core and steep power-law decay is fast, i.e.\ when $\eta$ in
  equation~(\ref{eq:rho}) is larger than $\sim2$.} are generally thought to be
unstable \citep[][\S5.5]{BinneyTremaine2008}. Hence, it seems unlikely that
such a model emerges from any reasonable formation mechanism. However, our
$N$-body models appears to be stable, and some further research is needed to
clarify the cause for `dynamical buoyancy', its relation to the core stalling
effect \citep[reported by][]{ReadEtAl2006a}, and its prospects in natural
systems.

For cusped mass models (IC \& SC), such as predicted by CDM cosmogony via
dissipationless simulations, clusters GC3 or GC4 will sink into the centre of
Fornax within 1-2\,Gyr with $\sim90\%$ probability (in the sense that
solutions where this does not occur cover only $\sim10\%$ of the possible
orbital phases and projections). In fact, we estimate the probability that no
cluster obtains $\rapo\lesssim100\,$pc within 2\,Gyr to be no more than 2\% in
Table~\ref{tab:prob:sinking}. This is largely independent of the assumed
globular-cluster orbital distribution function and constitues the more severe
\emph{immediate} timing problem in the sense of hypothesis \ref{hyp:short}
from page~\pageref{hyp:long}. It implies that \emph{if} Fornax indeed has a
cusped density profile, our cosmic epoch of observation is necessarily very
special.

For a shallow cusp model (WC), only the most massive cluster GC3 has a 90\%
probability to reach, within 1--2\,Gyr, $\rapo\lesssim200\,$pc significantly
less than its current projected radius of $\Robs=430\,$pc.  For this model, no
cluster can reach $\rapo\lesssim100\,$pc within 2\,Gyr and the chances that no
cluster will reach $\rapo\lesssim200\,$pc within 1 or 2\,Gyr are,
respectively, 92\% and 32\% (see Table~\ref{tab:prob:sinking}). These numbers
are not unlikely and hence avoid the long-term timing problem.
%%%%%%%%%%%%%%%%%%%%%%%%%%%%%%%%%%%%%%%%%%%%%%%%%%%%%%%%%%%%%%%%%%%%%%%%%%%%%%%%
\subsubsection{A steady-state solution}
Our finding thus suggest two possible solutions to the timing problem. First,
Fornax has a large core (perhaps between models LC and WC) and dynamical
friction is slow or has stalled a long time ago. In this case, Fornax may have
been on its current orbit for a Hubble time with its globular cluster system
hardly evolving. In this case, the consistency of the cluster distribution
with the stellar distribution (as discussed in \S\ref{sec:fornax:GCs}) cannot
be a coincidence, but hints at a common formation scenario.

%%%%%%%%%%%%%%%%%%%%%%%%%%%%%%%%%%%%%%%%%%%%%%%%%%%%%%%%%%%%%%%%%%%%%%%%%%%%%%%%
\subsubsection{An evolving solution}
In the second solution Fornax has a small core or shallow cusp (as model WC)
and dynamical friction is still ongoing, albeit slowly enough that the absence
of a central nucleus in Fornax (or in fact any other dSph) is perfectly
plausible. In this case, the clusters must have been further away from Fornax
in the past than today, and the current (weak) consistency of their
distribution with that of the stars is just a coincidence. Also, a Hubble time
ago the clusters most likely were more than the current tidal radius of
1.8--2.8\,kpc away from Fornax. This in turn requires Fornax did not orbit the
Milky Way for a Hubble time on its present orbit. However, even a simple
adiabatic evolution of Fornax's orbit may be sufficient to solve this
problem. For example, for a slowly growing Galaxy with always flat rotation
curve the peri-centric tidal radius of Fornax evolves like
$r_{\mathrm{tid}}^{}\propto\upsilon_{\mathrm{circ}}^{-4/3}$ (even when
neglecting mass-loss of Fornax due to Galactic tides).

This second solution appears more natural and also fits with the weak
indication of mass-segregation, as would be induced by dynamical friction, in
the current mass-radius relation (see Fig.~\ref{fig:GC:R:M}). However, this
model implies that the globular clusters have not formed within Fornax, but
are most likely accreted. One may, of course, consider these two solutions as
the extreme ends of a continuity of solutions with various degrees of cusp
strengths and hence dynamical friction effects.

%%%%%%%%%%%%%%%%%%%%%%%%%%%%%%%%%%%%%%%%%%%%%%%%%%%%%%%%%%%%%%%%%%%%%%%%%%%%%%%%
\subsection{The case of GC1} \label{sec:GC1}
An interesting aspect relates to the cluster GC1.
\cite*{PenarrubiaWalkerGilmore2009} demonstrated that, uniquely of all
Fornax's globular clusters, GC1 would be tidally disrupted if it fell to the
centre of Fornax. While we find that GC1 does not sink to the centre of Fornax
in almost all of our models, this still leaves us with a puzzle. Why should
the one cluster vulnerable to tides be on an orbit where it would hardly ever
suffer disruption? For our steady-state solution to the timing problem above,
this puzzle can be resolved by the postulation that Fornax once had a richer
globular-cluster system and we only see the survivors. Such survivors are
either massive enough or on \corr{remote} orbits to avoid tidal
disruption. However, a high (initial) frequency of globular clusters (of
$\sim10^{4-5}\Msun$) appears rather implausible for a small galaxy like
Fornax.

For the evolving solution to the timing problem, on the other hand, the fact
that GC1 would be disrupted poses no problem at all. In this picture, low-mass
clusters, such as GC1, would not be dragged down much, and there is no need to
postulate a large early population of clusters.

%%%%%%%%%%%%%%%%%%%%%%%%%%%%%%%%%%%%%%%%%%%%%%%%%%%%%%%%%%%%%%%%%%%%%%%%%%%%%%%%
\section*{Acknowledgments}
Research in Theoretical Astrophysics at Leicester is supported by an STFC
rolling grant. JIR would like to acknowledge support from SNF grant
PP00P2\_128540/1.

This research used the ALICE High Performance Computing Facility at the
University of Leicester. Some resources on ALICE form part of the DiRAC
Facility jointly funded by STFC and the Large Facilities Capital Fund of BIS.

%%%%%%%%%%%%%%%%%%%%%%%%%%%%%%%%%%%%%%%%%%%%%%%%%%%%%%%%%%%%%%%%%%%%%%%%%%%%%%%%
%\bibliographystyle{mn2e}
%\bibliography{paper,WDrefs}

%%%%%%%%%%%%%%%%%%%%%%%%%%%%%%%%%%%%%%%%%%%%%%%%%%%%%%%%%%%%%%%%%%%%%%%%%%%%%%%%
\appendix

%%%%%%%%%%%%%%%%%%%%%%%%%%%%%%%%%%%%%%%%%%%%%%%%%%%%%%%%%%%%%%%%%%%%%%%%%%%%%%%%
\section{Numerical convergence}
\label{app:converge}
In order to ensure that our simulations do not suffer from numerical noise we
ran simulations with two different mass models, one cusped and one cored; each
with two different orbits, one circular and one eccentric; and each of these
at three different resolutions: with $N=4\times10^5$, $10^6$ and
$4\times10^6$. We then compared the evolution in each case of one cluster over
10 Gyr.

\begin{figure}
  \centerline{
    \resizebox{78mm}{!}{\includegraphics{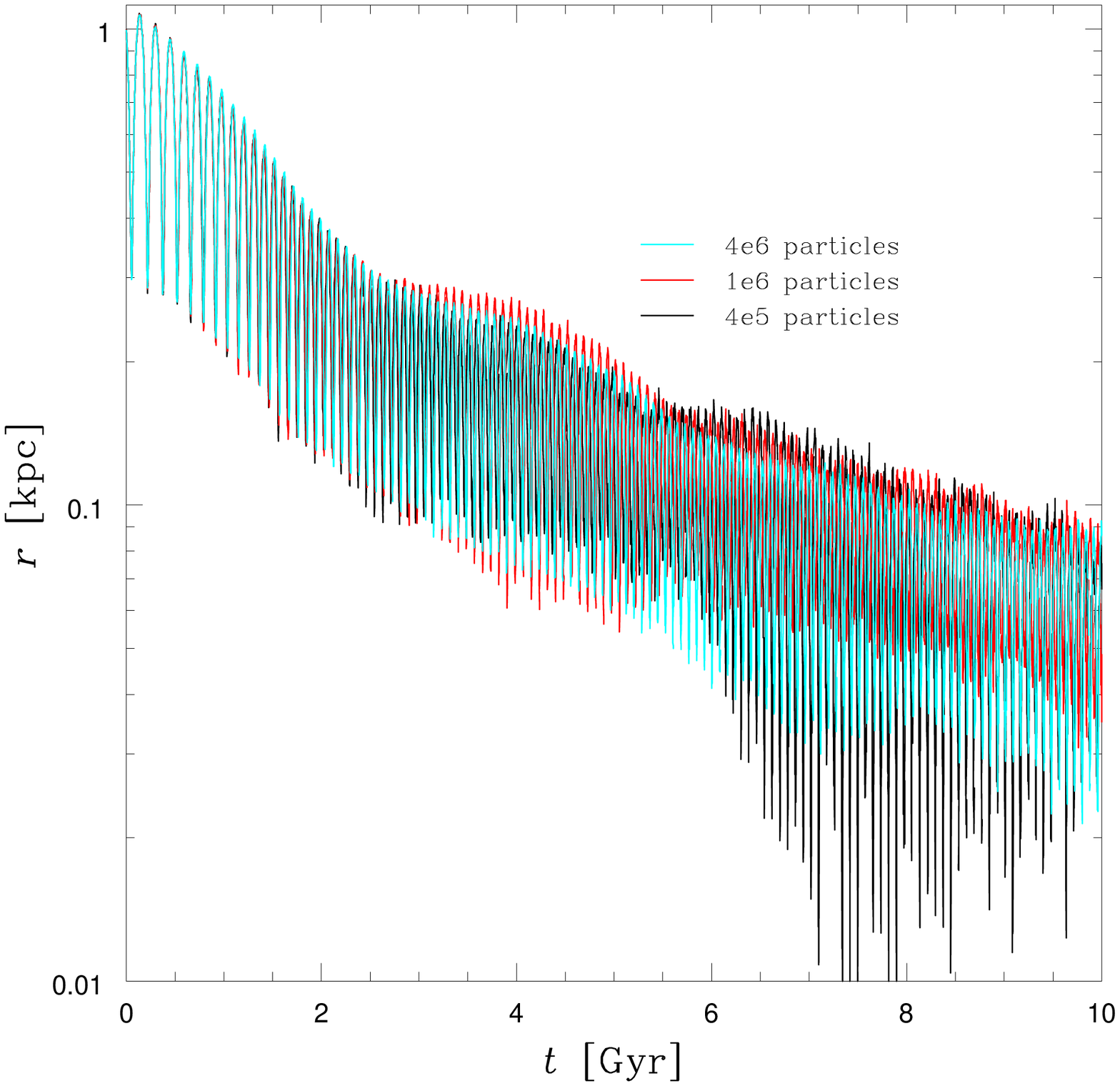}}
  }
  \caption{\label{concore2} The evolution of the radius for a single cluster
    moving on an eccentric orbit in a halo with a near-cored density profile
    (model WC from Table~\ref{tab:mass:models}), but realised with different
    particle numbers as indicated. Some of the deviation at later times is due
    to the uncertainties in determining the centre of the $N$-body model.}
\end{figure}

The evolution of the orbital radius of a single cluster moving on an eccentric
orbit in model WC (see table \ref{tab:mass:models}) is shown in
Fig.~\ref{concore2}. It can be seen that orbital evolution is very similar for
all resolutions. In particular, the decay of the orbit follows the same
timescale, with the time and radius of the first-stalling of the cluster being
the same. It has been shown that the two-body noise in a simulation can cause
the cluster orbit to precess and cause artificial decay of the orbit once in
the core \citep{ReadEtAl2006a}. We selected the above combination of orbit and
density profile precisely because \cite{ReadEtAl2006a} showed that convergence
is most difficult for an eccentric orbit in a cored halo. This is the case
where numerical friction caused by orbit precession has the largest effect on
orbital decay. The simulations shown above give a strong indication that such
effects are not significant at even lower resolutions than the one used for
the main body of this work. Our simulations are well converged.
%%
%%%%%%%%%%%%%%%%%%%%%%%%%%%%%%%%%%%%%%%%%%%%%%%%%%%%%%%%%%%%%%%%%%%%%%%%%%%%%%%%
\section{The probability for \boldmath $R\le\Robs$}
\label{app:prob}
The fraction of orbital phases and projections for which $R\le\Robs$ on a given
orbit is identical to the probability for $R\le\Robs$, when $R$ is computed
for that orbit at random orbital phase and projection.

The probability for a cluster with orbital energy $E$ and angular momentum $L$
to be found at a radius between $r$ and $r+d r$ is
\begin{equation}
  p(r|E,L) \, \diff r = \frac{2\,\diff r}{T_r\,\upsilon_r}
  \qquad\qquad\text{if}\; \rperi\le r\le \rapo
\end{equation}
and zero otherwise. $\rperi\le\rapo$ are the roots
of the radial velocity
\begin{equation}
  \upsilon_r^2 = 2\left[E-\Phi(r)\right]-L^2/r^2,
\end{equation}
and the radial period is given by
\begin{equation}
  T_r = 2 \int_{\rperi}^{\rapo}
  \frac{\diff r}{\upsilon_r}.
\end{equation}

The probability that a cluster at radius $r$ has projected radius between $R$
and $R+\diff R$ is
\begin{equation} \label{eq:pR}
  p(R|r)\,\diff R = \frac{R\,\diff R}{r\sqrt{r^2-R^2}} \qquad\qquad\text{if}\;
  R\le r
\end{equation}
and zero otherwise\footnote{This follows from the probability density
  $p(\theta)\diff\theta=\sin\theta$ for the polar angle $\theta\in[0,\pi]$
  and $p(R|r)\diff R=p(\theta)(\diff\theta/\diff R)\diff R$ with
  $R=r\sin\theta$.}. Thus, the probability for a cluster with given
orbit to have projected radius between $R$ and $R+\diff R$ is
\begin{equation}
  p(R|E,L)\,\diff R = \diff R \int_{\max\{R,\rperi\}}^{\rapo}
  p(R|r)\,p(r|E,L)\,\diff r
\end{equation}
for $R\le \rapo$ and zero otherwise. From this, we can work out the finite
probability that a cluster with given orbit has projected radius not greater
than observed:
\begin{eqnarray}
  p(R \le \Robs|E,L) &=&
  \int_0^{\Robs} p(R|E,L)\,\diff R
  \nonumber \\ &=& \frac{2}{T_r}
  \int_0^{\Robs} \diff R
  \int_{\max\{R,\rperi\}}^{\rapo}
  \frac{R}{r\sqrt{r^2-R^2}}
  \frac{\diff r}{\upsilon_r}
  \nonumber \\ &=& \frac{2}{T_r}
  \int_{\rperi}^{\rapo}\frac{\diff r}{\upsilon_r}
  \int_0^{\min\{r,\Robs\}} \frac{R\diff R}{r\sqrt{r^2-R^2}}
  \nonumber \\ &=& 
  \label{eq:plR}
  \int_{\rperi}^{\rapo}
  \left[1-(1-\Robs^2/r^2)_+^{1/2}\right]
  \frac{\diff r}{\upsilon_r}
  \Bigg/
  \int_{\rperi}^{\rapo}
  \frac{\diff r}{\upsilon_r}.
\end{eqnarray}
with $(\cdot)_+\equiv\max\{0,\cdot\}$. One can easily verify that
$p(R\le0|E,L)=0$ and $p(R\le\rapo|E,L)=1$, as required.

\label{lastpage}
\end{document}